\documentclass[11pt]{article}

\usepackage{amssymb}
\usepackage{amsmath,amsfonts,amsbsy,graphicx}
\usepackage{multirow}

\def\pr{{\partial}}
\def\eps{{\epsilon}}
\def\bx{{\bf x}}
\def\br{{\bf r}}
\def\bv{{\bf v}}
\def\bu{{\bf u}}
\def\bV{{\bf V}}
\def\be{{\bf e}}
\def\ba{{\bf a}}
\def\bc{{\bf c}}

\def\bbb{{\bf b}}

\def\bF{{\bf F}}

\def\bH{{\bf H}}
\def\bU{{\bf U}}
\def\bom{\boldsymbol{\omega}}

\def\bzeta{\boldsymbol{\zeta}}
\def\bsig{\boldsymbol{\sigma}}

\begin{document}


\centerline{\LARGE\bf Steady streaming between two vibrating planes}

\vskip 2mm
\centerline{\LARGE\bf at high Reynolds numbers}

\vskip 5mm
\centerline{\Large\bf {Konstantin Ilin}
\footnote{Department of Mathematics, University of York,
Heslington, York, YO10 5DD, U.K.. Electronic mail: konstantin.ilin@york.ac.uk}
and Andrey Morgulis
\footnote{Department of Mathematics, Mechanics and Computer Science, The Southern Federal University,
Rostov-on-Don, and South Mathematical Institute, Vladikavkaz Center of RAS, Vladikavkaz,
Russian Federation. Electronic mail: amor@math.rsu.ru}}

\vskip 2mm

\begin{abstract}
We consider incompressible flows between two transversely vibrating solid walls
and construct an asymptotic expansion of solutions of the Navier-Stokes equations
in the limit when both the amplitude of vibrations and the thickness of the Stokes layer are small and have the same order
of magnitude. Our asymptotic expansion is valid up to the flow boundary. In particular, we derive equations and boundary conditions, for the averaged flow.
In the leading order, the averaged flow is described by the stationary Navier-Stokes equations
with an additional term which contains the leading-order Stokes drift velocity. In a slightly different context (for a flow
induced by an oscillating conservative body force), the same equations
had been derived earlier by Riley \cite{Riley2001}.
The general theory is applied to two particular examples of steady streaming induced by transverse
vibrations of the walls in the form of standing and travelling plane waves. In particular, in the case of
waves travelling in the same direction, the induced flow is plane-parallel and the Lagrangian velocity profile can be computed analytically.
This example may be viewed as an extension of the theory of peristaltic pumping to the case of high Reynolds numbers.

\end{abstract}







\section{Introduction}
\label{Sec1}

\noindent
In this paper we study oscillating flows of a
viscous incompressible fluid between two solid walls produced by the transverse vibrations of the walls.
It is well-known that high-frequency
oscillations of the boundary of a domain occupied by a viscous
fluid generate not only an oscillating flow but also a (relatively) weak steady flow, which is usually
called the steady streaming (see, e.g., the review papers \cite{Lighthill} and \cite{Riley1967,Riley2001}).
Recently such flows attracted considerable attention
in the context of application of steady streaming to
micro-mixing \cite{Selverov, Yi, Carlsson} and to drag reduction in channel flows \cite{Hoepffner}.

\vskip 3mm
\noindent
The basic parameters of the problem
are the inverse Strouhal number $\epsilon$ and the
streaming Reynolds number $R_{s}$, defined as
\begin{equation}
\epsilon=\frac{V_{0}^{*}}{\omega d}, \quad
R_{s}=\frac{V_{0}^{*2}}{\omega \nu^{*}}
 \label{1.1}
\end{equation}
where $V_{0}^{*}$ is the amplitude of the velocity of the
oscillating walls, $d$ is the mean distance between the walls, $\omega$ is the angular
frequency of the vibrations and $\nu^{*}$ is the kinematic
viscosity of the fluid. Parameter $\epsilon$ measures the ratio of the amplitude of the
displacement of the vibrating wall to the mean distance between the walls and is
assumed to be small: $\epsilon \ll 1$. Note that the relation between
the standard Reynolds number $R$ based on $V_{0}^{*}$ and $d$
and the streaming Reynolds number $R_{s}$ is given by $R=V_{0}^{*}d/\nu^{*}=R_{s}/\epsilon$,
so that $R_s\sim 1$ corresponds to $R\gg 1$ when $\eps$ is small. The present paper deals with
the flow regimes with $R_s\sim 1$. This implies that
the amplitude of the wall displacement is of the same order of magnitude as the thickness
of the Stokes layer.

\vskip 3mm
\noindent
Steady streaming at $R_s\sim 1$ (and even
for large $R_s$) induced by translational oscillations of a rigid body
in a viscous fluid had been studied
by many authors (see \cite{Stuart1966, Riley1965, Wang1968, Bertelsen1973, Riley1975, HaddonRiley1979, Duck1979}).
In the case of a body in an infinite fluid, the problem can be reduced to an equivalent problem
about a fixed body placed in an oscillating flow, which considerably simplifies the analysis.
A related problem of steady streaming and mass transport produced by water waves had been treated in
\cite{Longuet1953, Longuet1983}) where flow regimes with large $R_s$ were considered.
Previous studies of flows produced by transverse oscillations of solid walls had been mostly focused
on the problem of peristaltic pumping in channels and pipes under the assumption
of low Reynolds numbers ($R\ll 1$) and small amplitude-to-wavelength ratio
(see, e.g., \cite{Jaffrin, Wilson, Carlsson}). Although there are a few paper where
the case of large Reynolds numbers ($R\gg 1$) had been considered \cite{Selverov, Yi, Hoepffner},
these studies still correspond to $R_{s}\ll 1$. As far as we know, the flow regimes with $R_s\sim 1$
have not been treated before.

\vskip 3mm
\noindent
The aim of this paper is to construct
an asymptotic expansion of the solution of the Navier-Stokes equation
in the limit $\eps\to 0$ and
$R_{s}\sim 1$. The procedure includes the derivation of the equations
and boundary conditions for a steady component of the flow (steady streaming) that
persists everywhere except thin layers near the vibrating walls
(whose thickness is of the same order as the amplitude of vibrations).
Since the thickness of the Stokes layer $\delta=\sqrt{\nu^*/\omega}$ and
the amplitude of the displacements of the walls $a$ are of the same order of magnitude,
the boundary conditions on the moving walls cannot be transferred
to the fixed mean positions of the walls. This is what makes the problem difficult.

\vskip 3mm
\noindent
To obtain
an asymptotic expansion,
we employ the Vishik-Lyusternik
method (see, e.g., \cite{Trenogin,
Nayfeh}) rather than the standard method of matched asymptotic expansions.
In comparison with the latter, the
Vishik-Lyusternik method does not
require the procedure of matching the inner and outer expansions
and the boundary layer part of the expansion
satisfies the condition of decay at infinity (in boundary layer variable)
in all orders of the expansion (this is not so in
the method of matched asymptotic expansions where the boundary layer part
usually does not decay and may even grow at infinity).
The Vishik-Lyusternik method had been used to study
viscous boundary layers at a fixed impermeable boundary by Chudov
\cite{Chud}. Recently, it has been applied to viscous boundary
layers in high Reynolds number flows through a fixed domain with
an inlet and an outlet \cite{Ilin2008}, to viscous flows in a
half-plane produced by tangential vibrations on its boundary
\cite{VV2008} and to the steady streaming between two cylinders \cite{Ilin2010}.
A similar technique had been used in \cite{Levenshtam} to construct an asymptotic expansion
in a problem of vibrational convection.

\vskip 3mm
\noindent
The outline of the paper is as follows. In Section 2, we formulate the mathematical
problem. In Section 3, the asymptotic equations and boundary conditions
are derived. Section 4 outlines the construction of the asymptotic solution.
In Section 5, we consider examples of the  steady streaming induced by vibrations of the walls
in the form of standing or travelling waves.
Finally, discussion of results and conclusions are presented in Section 6.


\setcounter{equation}{0}
\renewcommand{\theequation}{2.\arabic{equation}}

\section{Formulation of the problem}

\vskip 2mm
\noindent
We consider a three-dimensional viscous incompressible flow between two parallel walls
produced by  normal, periodic (in time) vibrations of the walls.
Let $\bx^{*}=(x^{*},y^{*},z^{*})$ be the Cartesian coordinates, $t^*$ the time,
$\bv^{*}=(u^{*},v^{*},w^{*})$ the velocity of the fluid, $p^{*}$ the pressure,
$\rho$ the fluid density, $\nu^{*}$ the kinematic viscosity and
$d$ the mean distance between the walls. We introduce
the non-dimensional quantities
\[
\tau=\omega t^{*}, \quad \bx=\frac{\bx^{*}}{d}, \quad \bv=\frac{\bv^*}{\omega \, a}, \quad
p=\frac{p^{*}}{\rho  \, a  \, d  \, \omega^2},
\]
where $\omega$ and $a$ are the angular frequency and the amplitude of the vibrations.
In these variables, the Navier-Stokes equations take the form
\begin{equation}
\bv_{\tau}+\eps \, (\bv\cdot\nabla)\bv = -\nabla p+\eps^2\nu \, \nabla\bv,
\quad \nabla\cdot\bv=0 . \label{1}
\end{equation}
where
\[
\eps=\frac{a}{d}, \quad \nu=\frac{\nu^{*}}{\omega a^2}.
\]
Parameter $\eps$ is the ratio of the amplitude of vibrations to the mean distance between the walls,
$\nu=R_{s}^{-1}$ where
$R_{s}$ is the Reynolds number based on the amplitudes of the velocity
and the normal displacement of the vibrating boundary (the streaming Reynolds number).
The walls are described by the equations
\begin{equation}
z=\eps \, f(x,y,\tau) \quad {\rm and} \quad
z=1+\eps \, g(x,y,\tau), \label{2}
\end{equation}
where $f(x,y,\tau)$ and $g(x,y,\tau)$ are given functions that are $2\pi$-periodic in $\tau$ and have
zero mean value, i.e.
\begin{equation}
\bar{f}(x,y)\equiv\frac{1}{2\pi}\int\limits_{0}^{2\pi}
f(x,y,\tau)\, d\tau =0, \quad
\bar{g}(x,y)\equiv\frac{1}{2\pi}\int\limits_{0}^{2\pi}
g(x,y,\tau)\, d\tau =0. \label{3}
\end{equation}
We also assume that $f(x,y,\tau)$ and $g(x,y,\tau)$ are periodic both in $x$ and in $y$
with periods $L_{x}$ and $L_{y}$, respectively.
Boundary conditions for the velocity at the walls are the standard no-slip
conditions:
\begin{equation}
\bv\!\bigm\vert_{z=\eps f}=f_{\tau}(x,y,\tau) \, \be_{z} , \quad
\bv\!\bigm\vert_{z=1+\eps g}=g_{\tau}(x,y,\tau) \, \be_{z}. \label{4}
\end{equation}
The incompressibility of the fluid
implies that functions $f$
and $g$ must satisfy an additional condition which, in the periodic case, is
given by
\begin{equation}
\int\limits_{0}^{L_{x}}
\int\limits_{0}^{L_{y}}f(x,y,\tau)\, dx \, dy =
\int\limits_{0}^{L_{x}}
\int\limits_{0}^{L_{y}}g(x,y,\tau) \, dx \, dy. \label{5}
\end{equation}
The non-dimensional equations and boundary conditions depend only on two parameters $\eps$ and $\nu$.
In what follows we are interested in
the asymptotic behaviour of periodic solutions of Eqs. (\ref{1}), (\ref{4}) in the limit when
$\eps\to 0$ and $\nu=O(1)$. In other words, we consider the situation where the amplitude of the vibration
is small and of the same order as the thickness of the Stokes layer.

\vskip 2mm
\noindent
It is convenient to separate vertical and horizontal components of the velocity as follows:
\[
\bv=\bu+ w \, \be_{z}, \quad \bu=u \, \be_{x}+v \, \be_{y},
\]
i.e. $\bu$ is the projection of the velocity onto the horizontal plane (the $xy$ plane). We will also
use the notation: $\br=(x,y)$ and $\bx=(x,y,z)$.

\vskip 2mm
\noindent
We seek a solution of (\ref{1}), (\ref{4}) in the form
\begin{eqnarray}
&&\bu=\bu^{r}(\br,z,\tau,\eps)+\bu^{a}(\br,\xi,\tau,\eps)+\bu^{b}(\br,\eta,\tau,\eps), \nonumber  \\
&&w=w^{r}(\br,z,\tau,\eps)+\eps \, w^{a}(\br,\xi,\tau,\eps)+\eps \, w^{b}(\br,\eta,\tau,\eps), \nonumber  \\
&&p=p^{r}(\br,z,\tau,\eps)+ p^{a}(\br,\xi,\tau,\eps)+ p^{b}(\br,\eta,\tau,\eps). \label{6}
\end{eqnarray}
Here $\xi=z/\epsilon$ and $\eta=(1-z)/\epsilon$ are the boundary layer variables;
all functions are assumed to be periodic in $\tau$, $x$ and $y$ with periods
$2\pi$, $L_{1}$ and $L_{2}$, respectively.
Functions $\bu^{r}$, $w^{r}$, $p^{r}$ represent
a regular expansion of the solution in power series in $\eps$ (an outer solution), and
$\{\bu^{a}, w^{a}, p^{a}\}$ and $\{\bu^{b}, w^{b}, p^{b}\}$  correspond to boundary layer corrections (inner solutions)
to this regular expansion.
We assume that the boundary layer parts of the expansion rapidly decay outside thin boundary layers,
namely:
\begin{equation}
\bu^{a}, w^{a}, p^{a} =o(\xi^{-s}) \quad {\rm as} \quad \xi\to\infty \quad {\rm and} \quad
\quad \bu^{b}, w^{b}, p^{b} =o(\eta^{-s}) \quad {\rm as} \quad \eta\to\infty    \label{7}
\end{equation}
for every $s>0$. In other words, we require that at  a distance of order unity from the wall,
the boundary layer corrections are smaller than any power of $\eps$.
This assumption will be verified {\it a posteriori}. We begin with the regular part of the expansion.


\setcounter{equation}{0}
\renewcommand{\theequation}{3.\arabic{equation}}

\section{Asymptotic expansion}

\subsection{Regular part of the expansion}

\noindent
Let
\begin{equation}
\bv^{r}=\bv^{r}_{0}+ \eps \, \bv^{r}_{1}+\eps^2 \bv^{r}_{2}+ \dots , \quad
p^{r}=p^{r}_{0}+ \eps \, p^{r}_{1}+\eps^2 p^{r}_{2}+ \dots , \label{3.1}
\end{equation}
where $\bv^{r}=\bu^{r}+ w^{r}\, \be_{z}$ and $\bv^{r}=\bu^{r}_{k}+ w^{r}_{k}\, \be_{z}$ ($k=0,1,2,\dots$).
The successive approximations $\bv^{r}_{k}$, $p^{r}_{k}$ ($k=0,1,2,\dots$)
satisfy the equations:
\begin{equation}
\pr_{\tau}\bv^{r}_{k}=-\nabla p^{r}_{k}+\bF^{r}_{k}(\bx,\tau), \quad
\nabla\cdot\bv^{r}_{0}=0, \label{3.2}
\end{equation}
where $\bF^{r}_{0}(\bx,\tau)\equiv 0$, $\bF^{r}_{1}(\bx,\tau)= -(\bv^{r}_{0}\cdot\nabla)\bv^{r}_{0}$ and
\[
\bF^{r}_{k}(\bx,\tau)= -\sum_{j=0}^{k-1}(\bv^{r}_{j}\cdot\nabla)\bv^{r}_{k-j-1}
+\nu\nabla^{2} \bv^{r}_{k-2} \quad (k=2,3\dots).
\]
In what follows, we will use the following notation: for any $2\pi$-periodic function $f(\tau)$,
\begin{equation}
f(\tau)=\bar{f}+\tilde{f}(\tau), \quad
\bar{f}=\frac{1}{2\pi}\int\limits_{0}^{2\pi}f(\tau)d\tau \label{3.3}
\end{equation}
where $\bar{f}$ is the mean value of $f(\tau)$ and, by definition, $\tilde{f}(\tau)=f(\tau)-\bar{f}$
is the oscillating part of $f(\tau)$ that
has zero mean value.

\subsubsection{Leading-order equations}

\noindent
Consider Eqs. (\ref{3.2}) for $k=0$. We seek a
solution $\bv^{r}_{0}$ which is periodic in $\tau$.
It can be written as
\begin{equation}
\bv^{r}_{0}=\bar{\bv}^{r}_{0}+\tilde{\bv}_{0}, \quad \tilde{\bv}^{r}_{0}=\nabla\phi_{0} \label{3.4}
\end{equation}
where $\phi_{0}$ has zero mean value and is the solution of the boundary value problem
\begin{eqnarray}
&&\nabla^{2}\phi_{0}=0, \quad \phi_{0}(x+L_x,y,z)=\phi_{0}(x,y,z),
 \quad \phi_{0}(x,y+L_y,z)=\phi_{0}(x,y,z), \nonumber \\
&&\phi_{0z}\!\bigm\vert_{z=0}=f_{\tau}(x,y,\tau), \quad  \phi_{0z}\!\bigm\vert_{z=1}=g_{\tau}(x,y,\tau) . \label{3.5}
\end{eqnarray}
Boundary conditions for $\phi_{0z}$ at $z=0$ and $z=1$ will be justified later.
On averaging Eqs. (\ref{3.2}) for $k=1$,
we obtain
\begin{equation}
(\bar{\bv}^{r}_{0}\cdot\nabla)\bar{\bv}^{r}_{0}
+\overline{(\tilde{\bv}^{r}_{0}\cdot\nabla)\tilde{\bv}^{r}_{0}}
=-\nabla \bar{p}^{r}_{1}, \quad \nabla\cdot\bar{\bv}^{r}_{0}=0. \label{3.6}
\end{equation}
Since, according to (\ref{3.4}), $\tilde{\bv}^{r}_{0}$ is irrotational, we have:
$\overline{(\tilde{\bv}^{r}_{0}\cdot\nabla)\tilde{\bv}^{r}_{0}}=
\overline{(\nabla\phi_{0}\cdot\nabla)\nabla\phi_{0}}=\nabla\bigl(
\overline{\vert\nabla\phi_{0}\vert^2}/2\bigr)$.
Therefore, we can rewrite (\ref{3.6}) as
\begin{equation}
(\bar{\bv}^{r}_{0}\cdot\nabla)\bar{\bv}^{r}_{0}
=-\nabla \Pi_{0}, \quad \nabla\cdot\bar{\bv}^{r}_{0}=0, \label{3.7}
\end{equation}
where $\Pi_{0}=\bar{p}^{r}_{1}+\overline{\vert\nabla\phi_{0}\vert^2/2}$.
Equations (\ref{3.7}) represent the time-independent Euler equations that describe
steady flows of an inviscid incompressible fluid.
It will be shown later that boundary conditions for $\bar{\bv}^{r}_{0}$ are
\begin{equation}
\bar{\bv}^{r}_{0}\!\bigm\vert_{z=0}={\bf 0}, \quad
\bar{\bv}^{r}_{0}\!\bigm\vert_{z=1}={\bf 0}. \label{3.7a}
\end{equation}
Thus, $\bar{\bv}^{r}_{0}(\bx)$ is a steady solution of the Euler equations
(\ref{3.7}), which is periodic in $x$ and $y$ and subject to zero boundary condition
at $z=0$ and $z=1$. We choose zero solution:
\begin{equation}
\bar{\bv}^{r}_{0}\equiv {\bf 0}  \label{3.9}
\end{equation}
(evidently, it satisfies the Euler equations as well as the boundary conditions). This choice
implies that there is no steady streaming in the leading order of the expansion.

\subsubsection{First-order equations}

\noindent
Separating the oscillatory part of Eqs. (\ref{3.2}) for $k=1$, we find that
\begin{equation}
\pr_{\tau}\tilde{\bv}^{r}_{1}=-\nabla \tilde{p}^{r}_{1}-\widetilde{(\bv^{r}_{0}\cdot\nabla)\bv^{r}_{0}}, \quad
\nabla\cdot\tilde{\bv}^{r}_{1}=0, \label{3.8}
\end{equation}
Using (\ref{3.9}) and the fact that $\tilde{\bv}^{r}_{0}$ is irrotational, (\ref{3.8})
can be written in the form
\begin{equation}
\pr_{\tau}\tilde{\bv}^{r}_{1}=-\nabla Q_{1}, \quad
\nabla\cdot\tilde{\bv}^{r}_{1}=0, \label{3.10}
\end{equation}
where $Q_{1}=\tilde{p}^{r}_{1}+\widetilde{(\nabla\phi_{0})^{2}}/2$.
It follows from (\ref{3.10}) that
\begin{equation}
\bv^{r}_{1}=\bar{\bv}^{r}_{1}+\tilde{\bv}_{1}, \quad \tilde{\bv}^{r}_{1}=\nabla\phi_{1} \label{3.11}
\end{equation}
where $\phi_{1}$ has zero mean value and is the solution of the problem
\begin{eqnarray}
&&\nabla^{2}\phi_{1}=0, \quad \phi_{1}(x+L_x,y,z)=\phi_{1}(x,y,z),
 \quad \phi_{1}(x,y+L_y,z)=\phi_{1}(x,y,z), \nonumber \\
&&\phi_{1z}\!\bigm\vert_{z=0}=a_{1}(x,y,\tau), \quad  \phi_{1z}\!\bigm\vert_{z=1}=b_{1}(x,y,\tau), \label{3.12}
\end{eqnarray}
where functions $a_{1}(x,y,\tau)$ and $b_{1}(x,y,\tau)$ will be defined later.

\vskip 3mm
\noindent
On averaging equation for $\bv^{r}_{3}$ and using (\ref{3.4}) and (\ref{3.9}), we find that
\begin{equation}
(\bar{\bv}^{r}_{1}\cdot\nabla)\bar{\bv}^{r}_{1}=
-\nabla \Pi^{*}_{3}+\nu\nabla^{2} \bar{\bv}^{r}_{1}-
\overline{\tilde{\bom}^{r}_{2}\times\nabla\phi_{0}}, \quad
\nabla\cdot\bar{\bv}^{r}_{1}=0. \label{3.13}
\end{equation}
where $\Pi^{*}_{3}=\bar{p}^{r}_{3}+
\overline{(\nabla\phi_{1})^{2}}/2+ \overline{\tilde{\bv}^{r}_{2}\cdot\nabla\phi_{0}}$
and $\tilde{\bom}^{r}_{2}=\nabla\times\tilde{\bv}^{r}_{2}$.
It is clear from Eq. (\ref{3.13}) that in order to obtain a closed equation for $\bar{\bv}^{r}_{1}$,
we need to find $\tilde{\bv}^{r}_{2}$.

\subsubsection{Second-order equations}

\noindent
Consider Eqs. (\ref{3.2}) for $k=2$. Averaging yields
\[
0=-\nabla \bar{p}^{r}_{2}-
\overline{(\bv^{r}_{0}\cdot\nabla)\bv^{r}_{1}}-
\overline{(\bv^{r}_{0}\cdot\nabla)\bv^{r}_{1}}, \quad \nabla\cdot \bar{\bv}^{r}_{2}=0.
\]
In view of (\ref{3.4}), (\ref{3.9}) and (\ref{3.11}), the first of these reduces
to the equation $\nabla \left(\bar{p}^{r}_{2}+
\overline{\nabla\phi_{0}\cdot\nabla\phi_{1}}\right)=0$
that can be integrated to obtain
\begin{equation}
\bar{p}^{r}_{2}=-
\overline{\nabla\phi_{0}\cdot\nabla\phi_{1}} + {\rm const}. \label{3.14}
\end{equation}
The oscillatory part of (\ref{3.2}) for $k=2$ gives us the equations for $\tilde{\bv}^{r}_{2}$:
\begin{equation}
\tilde{\bv}^{r}_{2\tau}=-\nabla Q_{2}-
(\tilde{\bv}^{r}_{0}\cdot\nabla)\bar{\bv}^{r}_{1}-
(\bar{\bv}^{r}_{1}\cdot\nabla)\tilde{\bv}^{r}_{0}, \quad \nabla\cdot\tilde{\bv}^{r}_{2}=0, \label{3.15}
\end{equation}
where $Q_{2}=\tilde{p}^{r}_{2}+
\widetilde{\nabla\phi_{0}\cdot\nabla\phi_{1}}$, and
we have used (\ref{3.4}), (\ref{3.9}) and (\ref{3.11}).
Taking {\em curl} of the first equation and using (\ref{3.4}), we obtain
\begin{equation}
\tilde{\bom}^{r}_{2\tau}=\left[\tilde{\bv}_{0}^{r},\bar{\bom}^{r}_{1}\right] \label{3.16}
\end{equation}
where $\bar{\bom}^{r}_{1}=\nabla\times\bar{\bv}^{r}_{1}$
and where $\left[\ba,\bbb\right]=(\bbb\cdot\nabla)\ba-(\ba\cdot\nabla)\bbb=\nabla\times(\ba\times\bbb)$
for any divergence-free vector fields $\ba(\bx)$ and $\bbb(\bx)$.
Now let $\bzeta(\bx,\tau)$ be such that
\begin{equation}
\bzeta_{\tau}=\tilde{\bv}^{r}_{0}, \quad \bar{\bzeta}={\bf 0}. \label{3.17}
\end{equation}
Then, it follows from (\ref{3.16}) that
\begin{equation}
\tilde{\bom}^{r}_{2}=\left[\bzeta,\bar{\bom}^{r}_{1}\right]. \label{3.18}
\end{equation}

\subsubsection{A closed system of equations for $\bar{\bv}^{r}_{1}$}

\noindent
In view of (\ref{3.17}) and (\ref{3.18}), we have
\begin{equation}
-\overline{\tilde{\bom}^{r}_{2}\times\nabla\phi_{0}}=\overline{\bzeta_{\tau}\times\tilde{\bom}^{r}_{2}}=
\overline{\bzeta_{\tau}\times \left[\bzeta,\bar{\bom}^{r}_{1}\right]}=-
\overline{\bzeta\times \left[\bzeta_{\tau},\bar{\bom}^{r}_{1}\right]}=
\overline{\bzeta\times \left[\bar{\bom}^{r}_{1},\bzeta_{\tau}\right]}, \label{3.20}
\end{equation}
where we have used the fact that
$\overline{f'(\tau)g(\tau)}=-\overline{f(\tau)g'(\tau)}$ for
any periodic functions $f$ and $g$. It follows from (\ref{3.20}) that
$-\overline{\tilde{\bom}^{r}_{2}\times\nabla\phi_{0}}=\frac{1}{2}
\left(
\overline{\bzeta_{\tau}\times \left[\bzeta,\bar{\bom}^{r}_{1}\right]}+
\overline{\bzeta\times \left[\bar{\bom}^{r}_{1},\bzeta_{\tau}\right]}\right)$.
With the help of the identity
$\ba\times[\bbb,\bc]+\bc\times[\ba,\bbb]+\bbb\times[\bc,\ba]=\nabla\left(\ba\cdot(\bbb\times\bc)\right)$ (which is valid for any divergence-free
$\ba$, $\bbb$  and $\bc$), this can be simplified
to
\[
-\overline{\tilde{\bom}^{r}_{2}\times\nabla\phi_{0}}
=\frac{1}{2} \, \overline{\left[\bzeta_{\tau},\bzeta\right]} \times \bar{\bom}^{r}_{1}
+\nabla \left(\bar{\bom}^{r}_{1}\cdot\overline{(\bzeta_{\tau}\times\bzeta)} \right).
\]
Finally, substituting the last formula into Eq. (\ref{3.13}), we obtain
\begin{equation}
(\bar{\bv}^{r}_{1}\cdot\nabla)\bar{\bv}^{r}_{1}=
-\nabla \Pi_{3}+\nu\nabla^{2} \bar{\bv}^{r}_{1}+
\bV\times \bar{\bom}^{r}_{1}. \label{3.21}
\end{equation}
where $\Pi_{3}=\Pi^{*}_{3}-\bar{\bom}^{r}_{1}\cdot\overline{(\bzeta_{\tau}\times\bzeta)}$ and
\begin{equation}
\bV=\frac{1}{2}\overline{\left[\bzeta_{\tau},\bzeta\right]}. \label{3.22}
\end{equation}
When $\bV$ is zero, (\ref{3.21}) coincides
with the stationary Navier-Stokes equations. In the two-dimensional case,
(\ref{3.22}) reduces to the equation derived earlier in \cite{Riley2001}.
It was also noticed in \cite{Riley2001} that $\bV$ represents the Stokes drift velocity of fluid particles.


\subsection{Boundary layer equations}

\noindent
In this subsection we will obtain asymptotic equations that describe boundary layers
near the vibrating walls.

\vskip 2mm
\noindent
{\it Boundary layer at the bottom wall}. To derive boundary layer equations
near the bottom wall, we ignore $\bu^{b}$, $w^{b}$ and $p^{b}$, because they
are supposed to be small relative to any power of $\eps$ everywhere except a thin boundary layer near $y=1$,
and assume that
\begin{equation}
\bu=\bu^{r}_{0}+\bu^{a}_{0}+ \eps (\bu^{r}_{1}+\bu^{a}_{1})+ \dots , \quad
w=w^{r}_{0}+ \eps (w^{r}_{1}+w^{a}_{0})+ \dots , \quad
p=p^{r}_{0}+p^{a}_{0}+ \eps (p^{r}_{1}+p^{a}_{1})+ \dots \quad \label{3.23}
\end{equation}
We insert (\ref{3.23})
into Eq. (\ref{1})  and take into account that $\bu^{r}_{k}$, $w^{r}_{k}$, $p^{r}_{k}$
($k=0,1,\dots$)
satisfy the equations (\ref{3.2}). Then we make the change of variables
$z=\eps \, \xi$, expand every function of $\eps \, \xi$ in Taylor's series at $\eps=0$
and collect terms of the equal powers in $\eps$. This produces the following
sequence of equations:
\begin{equation}
\pr_{\tau}\bu^{a}_{k}+w^{r}_{0}\!\bigm\vert_{z=0}\pr_{\xi}\bu^{a}_{k}+\nabla_{\|} \, p^{a}_{k}
-\nu\, \pr_{\xi}^2\bu^{a}_{k}=\bF_{k}^{a}, \quad
\pr_{\xi}p^{a}_{k}=G_{k}^{a}, \quad \nabla_{\|}\cdot\bu^{a}_{k}+\pr_{\xi}w^{a}_{k}=0 \label{3.24}
\end{equation}
for $k=0,1,\dots$
Here $\nabla_{\|}=\be_{x}\, \pr_{x}+\be_{y}\, \pr_{y}$. Functions $\bF_{k}^{a}$ and $G_{k}^{a}$
are defined in terms of lower order approximations. In particular,
$\bF_{0}^{a}\equiv 0$, $G_{0}^{a}\equiv 0$,
$G_{1}^{a}\equiv 0$ and
\begin{equation}
\bF_{1}^{a}=-\left(w^{r}_{1}+\xi \, w^{r}_{0z}\right)\!\bigm\vert_{z=0}\bu^{a}_{0\xi}
-\left(\bu^{r}_{0}\cdot\nabla_{\|}\right)\!\bigm\vert_{z=0}\bu^{a}_{0}-
\left(\bu^{a}_{0}\cdot\nabla_{\|}\right)\bu^{r}_{0}\!\bigm\vert_{z=0}-
\left(\bu^{a}_{0}\cdot\nabla_{\|}\right)\bu^{a}_{0}- w^{a}_{0}\bu^{a}_{0\xi}. \label{3.25}
\end{equation}

\vskip 2mm
\noindent
{\it Boundary layer at the upper wall}.
A similar procedure leads to the following equations of the boundary layer near the upper wall:
\begin{eqnarray}
\pr_{\tau}\bu^{b}_{k}-w^{r}_{0}\!\bigm\vert_{z=1}\pr_{\eta}\bu^{b}_{k}+\nabla_{\|} \, p^{b}_{k}
-\nu\, \pr_{\eta}^2\bu^{b}_{k}=\bF_{k}^{b}, \quad
\pr_{\eta}p^{b}_{k}=G_{k}^{b}, \quad \nabla_{\|}\cdot\bu^{b}_{k}-\pr_{\eta}w^{b}_{k}=0 \label{3.26}
\end{eqnarray}
for $k=0,1\dots$, where
$\bF_{0}^{b}\equiv 0$, $G_{0}^{b}\equiv 0$,
$G_{1}^{b}\equiv 0$ and
\begin{eqnarray}
\bF_{1}^{b}= \left(w^{r}_{1}-\eta \, w^{r}_{0z}\right)\!\bigm\vert_{z=1}\bu^{b}_{0\eta}
-\left(\bu^{r}_{0}\cdot\nabla_{\|}\right)\!\bigm\vert_{z=1}\bu^{b}_{0}-
(\bu^{b}_{0}\cdot\nabla_{\|})\bu^{r}_{0}\!\bigm\vert_{z=1}-
(\bu^{b}_{0}\cdot\nabla_{\|})\bu^{b}_{0}+ w^{b}_{0}\bu^{b}_{0\eta}. \label{3.27}
\end{eqnarray}
In accordance with (\ref{7}), we require that
\begin{equation}
\bu^{a}_{k}, w^{a}_{k}, p^{a}_{k} =o(\xi^{-s}) \quad {\rm as} \quad \xi\to\infty \quad {\rm and} \quad
\quad \bu^{b}_{k}, w^{b}_{k}, p^{b}_{k} =o(\eta^{-s}) \quad {\rm as} \quad \eta\to\infty    \label{3.28}
\end{equation}
for every $s>0$ and for each $k=0,1,\dots$.

\subsection{Boundary conditions}

\vskip 2mm
\noindent
Now we substitute (\ref{3.23}) in
(\ref{4}), expand all functions corresponding to the outer flow in Taylor's series at $z=0$
and collect terms of equal powers in $\eps$. This leads to the following boundary conditions
at the bottom wall:
\begin{eqnarray}
&&\bu^{r}_{0}\!\bigm\vert_{z=0}+ \, \bu^{a}_{0}\!\bigm\vert_{\xi=f}=0, \quad
w^{r}_{0}\!\bigm\vert_{z=0}=f_{\tau}; \label{3.29} \\
&&\bu^{r}_{k}\!\bigm\vert_{z=0}+
\bu^{a}_{k}\!\bigm\vert_{\xi=f}=- \sum_{n=1}^{k} \frac{f^n}{n!} \, \pr_{z}^n \bu^{r}_{k-n}\!\bigm\vert_{z=0}
\quad (k=1,2,\dots), \label{3.30} \\
&&w^{r}_{k}\!\bigm\vert_{z=0}= - w^{a}_{k-1}\!\bigm\vert_{\xi=f}
-\sum_{n=1}^{k} \frac{f^n}{n!} \, \pr_{z}^n w^{r}_{k-n}\!\bigm\vert_{z=0}\quad (k=1,2,\dots). \label{3.31}
\end{eqnarray}
A similar procedure yields the following boundary conditions
at the upper wall:
\begin{eqnarray}
&&\bu^{r}_{0}\!\bigm\vert_{z=1}+ \, \bu^{b}_{0}\!\bigm\vert_{\eta=-g}=0, \quad
w^{r}_{0}\!\bigm\vert_{z=1}=g_{\tau}; \label{3.32} \\
&&\bu^{r}_{k}\!\bigm\vert_{z=1}+
\bu^{b}_{k}\!\bigm\vert_{\eta=-g}=- \sum_{n=1}^{k} \frac{g^n}{n!} \, \pr_{z}^n \bu^{r}_{k-n}\!\bigm\vert_{z=1}
\quad (k=1,2,\dots), \label{3.33} \\
&&w^{r}_{k}\!\bigm\vert_{z=1}= - w^{b}_{k-1}\!\bigm\vert_{\eta=-g}
-\sum_{n=1}^{k} \frac{g^n}{n!} \, \pr_{z}^n w^{r}_{k-n}\!\bigm\vert_{z=1}\quad (k=1,2,\dots). \label{3.34}
\end{eqnarray}
Note that the boundary conditions for $\phi_{0z}$ at $z=0$  and $z=1$ in problem (\ref{3.5})
follow directly from (\ref{3.29}) and (\ref{3.32}).


\setcounter{equation}{0}
\renewcommand{\theequation}{4.\arabic{equation}}

\section{Construction of the asymptotic solution}

\subsection{Leading order equations}

\noindent
{\it Oscillatory outer flow}. In the leading order, the oscillatory outer flow
is irrotational, and the velocity potential $\phi_{0}$ is a solution
the boundary value problem (\ref{3.5}) (which is unique up to a constant).


\vskip 2mm
\noindent
{\it Boundary layer at the bottom wall}. In the leading order, the boundary layer is described
by Eqs. (\ref{3.24}) with $k=0$.
The decay condition (\ref{3.28}) for $p^{a}_{0}$ and the second equation (\ref{3.24})
imply that $p^{a}_{0}\equiv 0$. Hence, the equation for $\bu^{a}_{0}$
reduces to
\begin{equation}
\bu^{a}_{0\tau}+f_{\tau}\bu^{a}_{0\xi}=\nu \, \bu^{a}_{0\xi\xi}. \label{4.1}
\end{equation}
Here we have used boundary condition (\ref{3.29}) for $w^{r}_{0}$.
Let
\[
s=\xi-f(\br,\tau) \quad {\rm and} \quad \bu^{a}_{0}(\br,\xi,\tau)=\bU^{a}_{0}(\br,s(\br,\xi,\tau),\tau).
\]
Then Eq. (\ref{4.1}) simplifies to the standard heat equation for $\bU^{a}_{0}(\br,s,\tau)$:
\begin{equation}
\bU^{a}_{0\tau}=\nu \, \bU^{a}_{0ss}. \label{4.2}
\end{equation}
Boundary conditions for $\bU^{a}_{0}$ are the decay condition at infinity and
the condition
\begin{equation}
\bU^{a}_{0}\!\bigm\vert_{s=0}=-\bu_{0}^{r}\!\bigm\vert_{z=0}, \label{4.3}
\end{equation}
which follows from (\ref{3.29}).
Averaging Eq. (\ref{4.2}), we obtain the equation $\bar{\bU}^{a}_{0ss}=0$.
The only solution of this equation that satisfies the decay condition at infinity is zero solution
$\bar{\bU}^{a}_{0}\equiv 0$. [Note that
this doesn't mean that $\bar{\bu}^{a}_{0}=0$, because $\bU^{a}_{0}$ is averaged for  fixed $s$, and
$\bu^{a}_{0}$ - for fixed $\xi$. This fact, however, is inessential for what follows.]

\vskip 2mm
\noindent
A periodic (in $\tau$) solution of Eq. (\ref{4.2}) satisfying the decay condition at infinity
and boundary condition (\ref{4.3}) and having zero mean can be found by standard methods
(see, e.g., \cite{TikhonovSamarskii}).
Once $\bU^{a}_{0}$ is found, the normal velocity $w^{a}_{0}$ is determined from the continuity equation
(the third equation (\ref{3.24})) that can be written as
\begin{equation}
\nabla_{\|}\cdot\bU^{a}_{0}-\bU^{a}_{0s}\cdot\nabla_{\|}f+ W^{a}_{0s}=0 \label{4.4}
\end{equation}
where function $W^{a}_{0}(\br,s,\tau)$ is defined by the relation
$w^{a}_{0}(\br,\xi,\tau)=W^{a}_{0}(\br,s(\br,\xi,\tau),\tau)$.
Integration of Eq. (\ref{4.4}) in $s$ yields
\begin{equation}
W^{a}_{0}(\br,s,\tau)=\bU^{a}_{0}\cdot\nabla_{\|}f+
\nabla_{\|}\cdot\int\limits_{s}^{\infty}\bU^{a}_{0}(\br,s',\tau) \, ds', \label{4.5}
\end{equation}
where the constant of integration was chosen so as to guarantee that $W^{a}_{0}\to 0$ as
$s\to\infty$.

\vskip 2mm
\noindent
Boundary condition (\ref{3.31}) for $k=1$ is
\begin{equation}
w^{r}_{1}\!\bigm\vert_{z=0}=- \, f \, w^{r}_{0z}\!\bigm\vert_{z=0} - \,
w^{a}_{0}\!\bigm\vert_{\xi=f} \quad {\rm or, \ equivalently,} \quad
w^{r}_{1}\!\bigm\vert_{z=0}=- \, f \, w^{r}_{0z}\!\bigm\vert_{z=0} - \,
W^{a}_{0}\!\bigm\vert_{s=0}. \label{4.6}
\end{equation}
This means that function $a_{1}(x,y,\tau)$ in the boundary value problem (\ref{3.12}) is
given by
\begin{equation}
a_{1}(x,\tau)=- \, \widetilde{f \, w^{r}_{0z}}\!\bigm\vert_{z=0} - \,
\tilde{W}^{a}_{0}\!\bigm\vert_{s=0}. \label{4.7}
\end{equation}


\vskip 2mm
\noindent
{\it Boundary layer at the upper wall}. Exactly the same analysis as in the case of the bottom wall
leads to the heat equation
\begin{equation}
\bU^{b}_{0\tau}=\nu \, \bU^{b}_{0qq}, \label{4.8}
\end{equation}
where $q$ and $\bU^{b}$ are defined as
\[
q=\eta+g(\br,\tau) \quad {\rm and} \quad \bu^{b}_{0}(\br,\eta,\tau)=\bU^{b}_{0}(\br,q(\br,\eta,\tau),\tau).
\]
Boundary conditions for $\bU^{b}_{0}$ are the decay condition at infinity and
the condition
\begin{equation}
\bU^{b}_{0}\!\bigm\vert_{q=0}=-\bu_{0}^{r}\!\bigm\vert_{z=1}, \label{4.9}
\end{equation}
that follows from (\ref{3.32}).
Equation (\ref{4.8}) has a consequence that $\bar{\bU}^{b}_{0qq}=0$.
This, together with the condition of decay at infinity (\ref{3.28}),
implies that $\bar{\bU}^{b}_{0}\equiv 0$.

\vskip 2mm
\noindent
Again, an oscillatory solution of Eq. (\ref{4.8}) can be found by standard methods.
The normal velocity $w^{b}_{0}$ is determined from the continuity equation
(the third equation (\ref{3.26})):
\begin{equation}
W^{b}_{0}(\br,q,\tau)=\bU^{b}_{0}\cdot\nabla_{\|}g
-\nabla_{\|}\cdot\int\limits_{q}^{\infty}\bU^{b}_{0}(\br,q',\tau) \, dq', \label{4.10}
\end{equation}
where function $W^{b}_{0}$ is defined by
$w^{b}_{0}(\br,\eta,\tau)=W^{b}_{0}(\br,q(\br,\eta,\tau),\tau)$.
and where the constant of integration was chosen so as to guarantee that $W^{b}_{0}\to 0$ as
$q\to\infty$.

\vskip 2mm
\noindent
Boundary condition (\ref{3.34}) for $k=0$ can be written as
\begin{equation}
w^{r}_{1}\!\bigm\vert_{z=1}=- \, g \, w^{r}_{0z}\!\bigm\vert_{z=1} - \,
W^{b}_{0}\!\bigm\vert_{q=0}. \label{4.11}
\end{equation}
This means that function $b_{1}(x,y,\tau)$ in the boundary value problem (\ref{3.12}) is
given by
\begin{equation}
b_{1}(x,\tau)=- \, \widetilde{g \, w^{r}_{0z}}\!\bigm\vert_{z=1} - \,
\tilde{W}^{b}_{0}\!\bigm\vert_{q=0}. \label{4.12}
\end{equation}


\vskip 2mm
\noindent
{\it Averaged outer flow}. Averaging
boundary conditions (\ref{3.29}) and (\ref{3.32}), we find that
\begin{eqnarray}
&&\bar{\bu}^{r}_{0}\!\bigm\vert_{z=0}=- \, \overline{\bu^{a}_{0}\!\bigm\vert_{\xi=f}}
=-\bar{\bU}^{a}_{0}\!\bigm\vert_{s=0}=0, \quad
\bar{w}^{r}_{0}\!\bigm\vert_{z=0}=\overline{f_{\tau}}=0,  \nonumber \\
&&\bar{\bu}^{r}_{0}\!\bigm\vert_{z=1}=- \, \overline{\bu^{b}_{0}\!\bigm\vert_{\eta=-g}}
=-\bar{\bU}^{b}_{0}\!\bigm\vert_{q=0}=0, \quad
\bar{w}^{r}_{0}\!\bigm\vert_{z=1}=\overline{g_{\tau}}=0.  \nonumber
\end{eqnarray}
This justifies our earlier assumption (\ref{3.7a}) and our conclusion (\ref{3.9}) that
{\it there is no steady streaming
in the leading order of the expansion}.

\subsection{First-order equations}

\noindent
{\it Oscillatory outer flow}. The oscillatory part of the (first-order) outer flow is determined
by the boundary value problem (\ref{3.12}) for the Laplace equation.
Since functions $a_{1}(\br,\tau)$ and $b_{1}(\br,\tau)$, which appear
in the boundary conditions for $\phi_{1z}$, are now known, problem (\ref{3.12}) can be solved, thus giving us
$\tilde{\bv}^{r}_{1}(\bx,\tau)$.


\vskip 3mm
\noindent
{\it Boundary layer at the bottom wall}.
Consider now the first-order boundary layer equations (Eqs. (\ref{3.24}) for $k=1$).
Again, the condition of decay at infinity for $p^{a}_{1}$ and the second equation (\ref{3.24})
imply that $p^{a}_{1}\equiv 0$, so that the first equation (\ref{3.24}) simplifies to
\begin{equation}
\bu^{a}_{1\tau}+ \, f_{\tau} \, \bu^{a}_{1\xi}-\nu \bu^{a}_{1\xi\xi}=\bF_{1}^{a}. \label{4.13}
\end{equation}
We are looking for a solution of (\ref{4.13}) which satisfies the decay condition at infinity
(\ref{3.28}) and boundary condition (\ref{3.30}) for $k=1$. To find it, we again employ variable $s=\xi-f(\br,\tau)$ and
rewrite (\ref{4.13}) and (\ref{3.30}) in the form
\begin{eqnarray}
&&\bU^{a}_{1\tau}-\nu \bU^{a}_{1ss}=\bH^{a}_{1},  \label{4.14} \\
&&\bU^{a}_{1}\!\bigm\vert_{s=0}=-\bu^{r}_{1}\!\bigm\vert_{z=0}- \, f \, \bu^{r}_{0z}\!\bigm\vert_{z=0}, \label{4.15}
\end{eqnarray}
where $\bU^{a}_{1}(\br,s,\tau)$ is such that $\bU_{1}(\br,s(\br,\xi,\tau),\tau)=\bu^{a}_{1}(\br,\xi,\tau)$
and where
\begin{eqnarray}
\bH^{a}_{1}&=&-\left(w^{r}_{1}+(s+f) \, w^{r}_{0z}\right)\!\bigm\vert_{z=0}\bU^{a}_{0s}
-\left(\bu^{r}_{0}\cdot\nabla\right)\!\bigm\vert_{z=0}\bU^{a}_{0}
+\left(\bu^{r}_{0}\cdot\nabla f\right)\!\bigm\vert_{z=0}\bU^{a}_{0s} \nonumber \\
&&- \left(\bU^{a}_{0}\cdot\nabla\right)\bu^{r}_{0}\!\bigm\vert_{z=0}-
\left(\bU^{a}_{0}\cdot\nabla\right)\bU^{a}_{0}
+\left(\bU^{a}_{0}\cdot\nabla f\right)\bU^{a}_{0s}
- W^{a}_{0} \, \bU^{a}_{0s}. \label{4.16}
\end{eqnarray}
Averaging (\ref{4.14}), we find that
\begin{equation}
\nu \bar{\bU}^{a}_{1ss}=-\bar{\bH}^{a}_{1}(\br,s). \label{4.17}
\end{equation}
We integrate this equation two times, choosing constants of integration so as
to satisfy the condition of decay at infinity in variable $s$. As a result, we have
\begin{equation}
\bar{\bU}^{a}_{1}=-\frac{1}{\nu}\int\limits_{s}^{\infty}\int\limits_{s'}^{\infty}
\bar{\bH}_{1}^{a}(\br,s'') \, ds'' \, ds'. \label{4.18}
\end{equation}
Note that (\ref{4.18}) is the unique solution of Eq. (\ref{4.17}) that decays at infinity.
The oscillatory part of $\bU^{a}_{1}$ can also be found from Eqs. (\ref{4.14}) and (\ref{4.15}),
but we will not do it here as we are only interested in the averaged part of the flow.


\vskip 3mm
\noindent
{\it Boundary layer at the upper wall}.
A similar analysis leads to the equations
\begin{eqnarray}
&&\bU^{b}_{1\tau}-\nu \bU^{b}_{1qq}=\bH^{b}_{1},  \label{4.19} \\
&&\bU^{b}_{1}\!\bigm\vert_{q=0}=-\bu^{r}_{1}\!\bigm\vert_{z=1}- \, g \, \bu^{r}_{0z}\!\bigm\vert_{z=1}, \label{4.20}
\end{eqnarray}
where $\bU^{b}_{1}(\br,q,\tau)$ is defined by the relation
$\bU_{1}^{b}(\br,q(\eta,\tau),\tau)=\bu^{b}_{1}(\br,\eta,\tau)$
and where
\begin{eqnarray}
\bH^{b}_{1}&=&\left(w^{r}_{1}-(q-g) \, w^{r}_{0z}\right)\!\bigm\vert_{z=1}\bU^{b}_{0q}
-\left(\bu^{r}_{0}\cdot\nabla\right)\!\bigm\vert_{z=1}\bU^{b}_{0}
-\left(\bu^{r}_{0}\cdot\nabla g\right)\!\bigm\vert_{z=1}\bU^{b}_{0q} \nonumber \\
&&- \left(\bU^{b}_{0}\cdot\nabla\right)\bu^{r}_{0}\!\bigm\vert_{z=1}-
\left(\bU^{b}_{0}\cdot\nabla\right)\bU^{b}_{0}-
\left(\bU^{b}_{0}\cdot\nabla g\right)\bU^{b}_{0q}
+ W^{b}_{0} \, \bU^{b}_{0q}. \label{4.21}
\end{eqnarray}
Averaging (\ref{4.19}), we obtain
\begin{equation}
\nu \bar{\bU}^{b}_{1qq}=-\bar{\bH}^{b}_{1}(\br,q). \label{4.22}
\end{equation}
Integration of (\ref{4.22}) yields
\begin{equation}
\bar{\bU}^{b}_{1}=-\frac{1}{\nu}\int\limits_{q}^{\infty}\int\limits_{q'}^{\infty}
\bar{\bH}^{b}_{1}(\br,q'') \, dq'' \, dq'. \label{4.23}
\end{equation}
Here again the constants of integration are chosen so as to satisfy the condition of decay at infinity.


\vskip 2mm
\noindent
{\it Averaged outer flow}.
Averaging the boundary conditions (\ref{3.30}) and (\ref{3.33}) for $k=1$,
we obtain
\begin{eqnarray}
&&\bar{\bu}^{r}_{1}\!\bigm\vert_{z=0}=-\bar{\bU}_{1}^{a}\!\bigm\vert_{s=0}
- \, \overline{f \, \bu^{r}_{0z}}\!\bigm\vert_{z=0}, \label{4.24} \\
&&\bar{\bu}^{r}_{1}\!\bigm\vert_{z=1}=-\bar{\bU}_{1}^{b}\!\bigm\vert_{q=0}
- \, \overline{g \, \bu^{r}_{0z}}\!\bigm\vert_{z=1}. \label{4.25}
\end{eqnarray}
Once $\bar{\bU}_{1}^{a}$ and $\bar{\bU}_{1}^{b}$ are found (from Eqs. (\ref{4.18})
and (\ref{4.23})), Eqs. (\ref{4.24}) and (\ref{4.25}) give
us boundary conditions for $\bar{\bu}^{r}_{1}$.

\vskip 2mm
\noindent
Boundary conditions for $\bar{w}^{r}_{1}$ are obtained by averaging
Eqs. (\ref{4.6}) and (\ref{4.11}):
\begin{eqnarray}
&&\bar{w}^{r}_{1}\!\bigm\vert_{z=0}=- \, \overline{f \, w^{r}_{0z}}\!\bigm\vert_{z=0} - \,
\overline{\bU^{a}_{0}\cdot\nabla f}\!\bigm\vert_{s=0}, \nonumber \\
&&\bar{w}^{r}_{1}\!\bigm\vert_{z=1}=- \, \overline{g \, w^{r}_{0z}}\!\bigm\vert_{z=1} - \,
\overline{\bU^{b}_{0}\cdot\nabla g}\!\bigm\vert_{q=0}. \nonumber
\end{eqnarray}
Here we have used Eqs. (\ref{4.5}) and (\ref{4.10}).
With the help of (\ref{4.3}) and (\ref{4.9}), these can be transformed to
\begin{eqnarray}
&&\bar{w}^{r}_{1}\!\bigm\vert_{z=0}=\nabla_{\|} \cdot
\overline{(f \nabla_{\|}\phi_{0})}\!\bigm\vert_{z=0}, \label{4.26} \\
&&\bar{w}^{r}_{1}\!\bigm\vert_{z=1}=\nabla_{\|} \cdot
\overline{(g \nabla_{\|}\phi_{0})}\!\bigm\vert_{z=1}. \label{4.27}
\end{eqnarray}
Equations (\ref{4.24})--(\ref{4.27})
represent boundary conditions for
the first-order
averaged equations for the outer flow (\ref{3.21}).

\vskip 2mm
\noindent
{\it Remark on boundary conditions} (\ref{4.26}), (\ref{4.27}). In general case the right sides
of Eqs. (\ref{4.26}) and (\ref{4.27}) are nonzero. Nevertheless, the net flux of the averaged
Eulerian velocity through any rectangle
of periods in $x$ and $y$ always vanishes due to periodicity of
$f$, $g$ and $\phi_{0}$ and the form of the right sides of Eqs. (\ref{4.26}), (\ref{4.27}). It can be shown
(see Appendix A) that the $z$-component of the Stokes drift velocity
at $z=0$ and $z=1$ is equal to minus
the right sides of Eqs. (\ref{4.26}) and (\ref{4.27}), so that
the averaged Lagrangian velocity at $z=0$ and $z=1$ vanishes. This means that
at each point of the boundary the averaged mass flux of the fluid is exactly zero.


\setcounter{equation}{0}
\renewcommand{\theequation}{5.\arabic{equation}}

\section{Examples}

\subsection{Example 1: standing waves}

\noindent
Let the vibrations of the walls be in the form of standing waves:
\begin{equation}
f(\br,\tau)=R(\br) \, h(\tau), \quad g(\br,\tau)=S(\br) \, h(\tau), \label{5.1}
\end{equation}
where $R(\br)$ and $S(\br)$ are given functions,  periodic in $x$ and $y$ with periods
$L_{x}$ and $L_{y}$, respectively, and satisfying condition (\ref{5}), and where
$h(\tau)$ is a given $2\pi$-periodic function. In this case, the solution of
problem (\ref{3.5}) has the form $\phi_{0}(\br,z,\tau)=\Phi(\br,z) \, h'(\tau)$
where $\Phi(\br,z)$ is the solution of the time-independent problem
\begin{eqnarray}
&&\nabla^2\Phi=0, \quad \Phi(x+L_{x},y,z)=\Phi(x,y,z),
\quad \Phi(x,y+L_{y},z)=\Phi(x,y,z),\nonumber \\
&&\Phi_{z}\!\bigm\vert_{z=0}=R(\br), \quad
\Phi_{z}\!\bigm\vert_{z=1}=S(\br). \label{5.2}
\end{eqnarray}
It is not difficult to show that
\[
\overline{f \, \bu^{r}_{0z}}\!\bigm\vert_{z=0}=0, \quad
\overline{g \, \bu^{r}_{0z}}\!\bigm\vert_{z=1}=0,  \quad
\nabla_{\|} \cdot
\overline{(f \nabla_{\|}\phi_{0})}\!\bigm\vert_{z=0}=0,  \quad
\nabla_{\|} \cdot
\overline{(g \nabla_{\|}\phi_{0})}\!\bigm\vert_{z=1}=0.
\]
Therefore,
boundary conditions (\ref{4.24})--(\ref{4.27}) simplify to
\begin{equation}
\bar{\bu}^{r}_{1}\!\bigm\vert_{z=0}=-\bar{\bU}_{1}^{a}\!\bigm\vert_{s=0}, \quad
\bar{\bu}^{r}_{1}\!\bigm\vert_{z=1}=-\bar{\bU}_{1}^{b}\!\bigm\vert_{q=0}, \quad
\bar{w}^{r}_{1}\!\bigm\vert_{z=0}=0, \quad
\bar{w}^{r}_{1}\!\bigm\vert_{z=1}=0. \label{5.3}
\end{equation}
Thus, vibrations of the walls in the form of standing waves produce zero boundary conditions
for the normal velocity in the averaged outer flow.

\vskip 2mm
\noindent
It is also easy to deduce from (\ref{3.22}) that
the Stokes drift velocity $\bV$ is zero, so that
Eq. (\ref{3.21}) reduces to
\[
\left(\bar{\bv}^{r}_{1}\cdot\nabla\right)\bar{\bv}^{r}_{1}=-\nabla\Pi_{3}+
\nu\nabla^{2} \bar{\bv}^{r}_{1}.
\]
Thus, in this example, the first-order steady outer flow is described by
the Navier-Stokes equations with boundary conditions (\ref{5.3}).

\begin{figure}
\begin{center}
\includegraphics*[height=8cm]{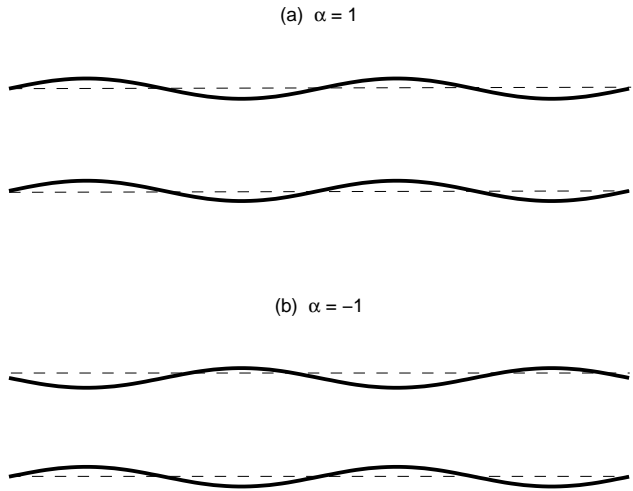}
\end{center}
\vskip -10mm
\caption{(a) bending waves; (b) contraction/expansion waves.}
\label{fig1}
\end{figure}

\vskip 2mm
\noindent
In particular, if we consider vibrations of the walls in the form of plane standing waves
\[
f(x,y,\tau)=\cos(k x)\cos\tau, \quad g(x,y,\tau)=\alpha\cos(k x)\cos\tau,
\]
where $k=2\pi/L_{x}$ is the wave number, $\alpha=\pm 1$ ($\alpha=1$ corresponds to bending waves
and $\alpha= - 1$ - to contraction/expansion waves, see Fig. \ref{fig1}), then
the boundary conditions (\ref{5.3}) will take the form
\begin{equation}
\bar{\bv}^{r}_{1}\!\bigm\vert_{z=0}=
-\frac{3}{8} \, k \, Q^2(k) \, \sin(2kx) \, \be_{x}, \quad
\bar{\bv}^{r}_{1}\!\bigm\vert_{z=1}=
-\frac{3}{8}  \, k \, Q^2(k) \, \sin(2kx) \, \be_{x}, \label{5.3a}
\end{equation}
where
\[
Q(k)=\frac{1-\alpha\cosh k}{\sinh k }.
\]

\begin{figure}
\begin{center}
\includegraphics*[height=7.5cm]{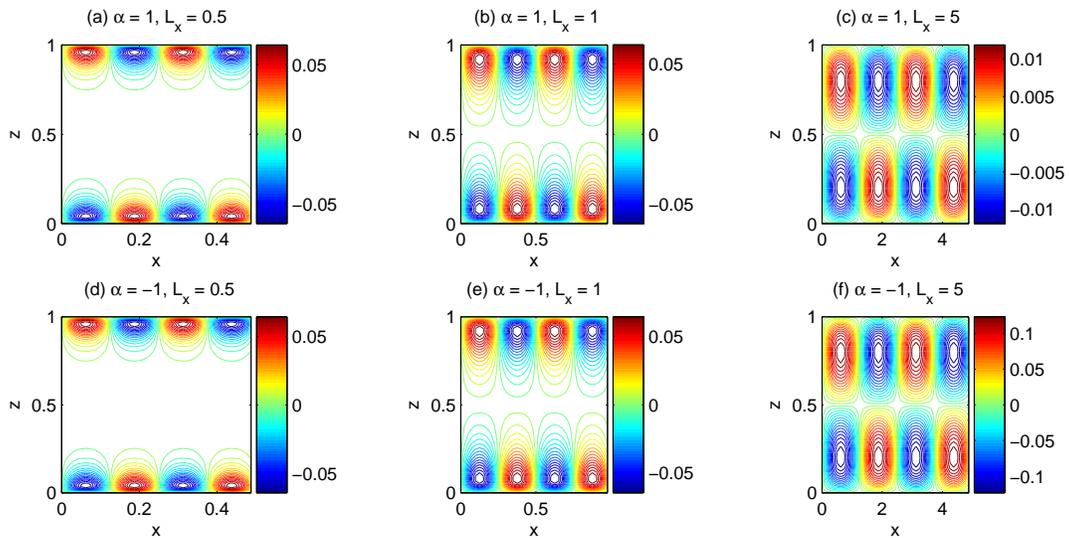}
\end{center}
\vskip -10mm
\caption{Streamlines of the averaged velocity for $\nu=1$ (standing waves).}
\label{fig1a}
\end{figure}

\vskip 2mm
\noindent
The results of numerical solution of the Navier-Stokes equations subject
to the boundary conditions (\ref{5.3a}) are shown in Fig. \ref{fig1a}.
Qualitatively, the stream line picture is the same for both bending and
contraction/expansion waves. The magnitudes of the averaged flow in these two cases are different
for $L_x>1$.
Let us measure the magnitude of the velocity field by
$\max\vert\bar{\bv}_{1}^{r}\vert$ where the maximum is computed for all
$(x,z)$ such that
$0\leq x\leq L_{x}$, $0\leq z\leq 1$. Table 1 shows $\max\vert\bar{\bv}_{1}^{r}\vert$
for various $L_{x}$.
Evidently, the maximum velocity decreases, when wavelength $L_{x}$ increases, much faster for bending waves ($\alpha=1$)
than for contraction/expansion waves ($\alpha=-1$).

\vskip -3mm
\begin{table}[h!]
\begin{center}
\begin{tabular}{|c|c|c|c|c|}
\hline
\multicolumn{2}{|c|}{$L_x$}  &0.5 &1 &5 \\
\hline
\multirow{2}{*}{$\max\vert\bar{\bv}_{1}^{r}\vert$ } &$\alpha=1$ &2.3562 &1.1693   &0.073073   \\
&$\alpha=-1$ &2.3562 &1.1869   &0.75974   \\
\hline
\end{tabular}
\end{center}
  \vskip -3mm
  \caption{$\max\vert\bar{\bv}_{1}^{r}\vert$ for standing waves and $\nu=1$.}
\end{table}

\subsection{Example 2: travelling waves}

\noindent
Consider now vibrations of the walls in the form of plane travelling waves:
\begin{equation}
f(\br,\tau)=\cos(kx-\tau), \quad g(\br,\tau)=\alpha \cos(kx-\beta\tau), \label{5.4}
\end{equation}
where $k=2\pi/L_{x}$, $\alpha=\pm 1$ and $\beta=\pm 1$. If $\beta=1$, both waves travel in the same direction,
if $\beta=-1$, the waves travel in opposite directions.
Since both $f$ and $g$ do not depend on
$y$, we will look for a solution which does not depend on $y$.
Problem (\ref{3.5}) simplifies to
\begin{eqnarray}
&&\phi_{0xx}+\phi_{0zz}=0, \quad \phi_{0}(x+L_{x},z)=\phi_{0}(x,z), \nonumber \\
&&\phi_{0z}\!\bigm\vert_{z=0}=\sin(kx-\tau), \quad
\phi_{0z}\!\bigm\vert_{z=1}=\alpha \sin(kx-\tau). \label{5.5}
\end{eqnarray}
Consider first $\beta=1$, i.e. both waves travel in the positive direction of the $x$ axis.
Boundary conditions (\ref{4.24})--(\ref{4.27}) become
\begin{equation}
\bar{u}_{1}^{r}\!\bigm\vert_{z=0}=A(k), \quad
\bar{u}_{1}^{r}\!\bigm\vert_{z=1}=A(k), \quad
\bar{w}_{1}^{r}\!\bigm\vert_{z=0}=0, \quad
\bar{w}_{1}^{r}\!\bigm\vert_{z=1}=0, \label{5.6}
\end{equation}
where
\[
A(k)=\frac{k}{4\sinh^2(k)}\left[5+\cosh^2(k)-6\alpha \cosh(k)\right]
\]
The Stokes drift velocity is given by
\begin{equation}
\bV(\br)=V_{0}(z)\be_{x}, \quad
V_{0}(z)=\frac{k\left[\cosh(k)-\alpha\right]}{\sinh^2(k)}\cosh\left[k(2z-1)\right] . \label{5.7}
\end{equation}
Now $\bar{\bom}_{1}^{r}=\Omega(x,z)\be_{y}$ where
$\Omega(x,z)=\pr_{z}\bar{u}_{1}^{r}-\pr_{x}\bar{w}_{1}^{r}$, and
Eq. (\ref{3.21}) can be written as
\begin{equation}
(\bar{\bv}^{r}_{1}\cdot\nabla)\bar{\bv}^{r}_{1}=
-\nabla \Pi_{3}+\nu\nabla^{2} \bar{\bv}^{r}_{1}+
V_0(z) \Omega \, \be_{z}. \label{5.8}
\end{equation}
Equation (\ref{5.8}) subject to boundary conditions (\ref{5.6}) has the constant solution
\begin{equation}
\bar{\bv}_{1}^{r}=A(k)\be_{x}. \label{5.9}
\end{equation}

\begin{figure}
\begin{center}
\includegraphics*[height=6cm]{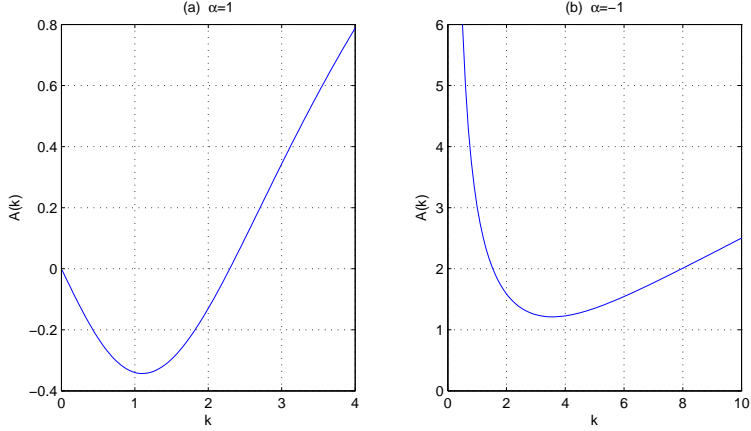}
\end{center}
\vskip -8mm
\caption{The averaged Eulerian velocity as function of $k$.}
\label{fig2}
\end{figure}

\noindent
The constant horizontal velocity as function of $k$ is shown in Fig. \ref{fig2}.
For bending waves, it is negative for small $k$ and positive for large $k$ ($A(k)\sim k$ for
both $k\ll 1$ and $k\gg 1$).
For contraction/expansion waves, it is always positive and growing both when $k\to 0$ and when $k\to\infty$
($A(k)\sim 1/k$ for $k\ll 1$ and $A(k)\sim k$ for $k\gg 1$).

\vskip 2mm
\noindent
Although Eq. (\ref{5.8}) also has solutions
with a nonzero pressure gradient $\nabla \Pi_{3}=c_{0}\be_{x}$ ($c_{0}=const$), we do not consider
such solutions here, because this would amount to a modification of our
original problem, allowing the presence of a weak $O(\eps^3)$ pressure gradient.
Note also that flows with stronger pressure gradients would require a separate treatment.

\begin{figure}
\begin{center}
\includegraphics*[height=6cm]{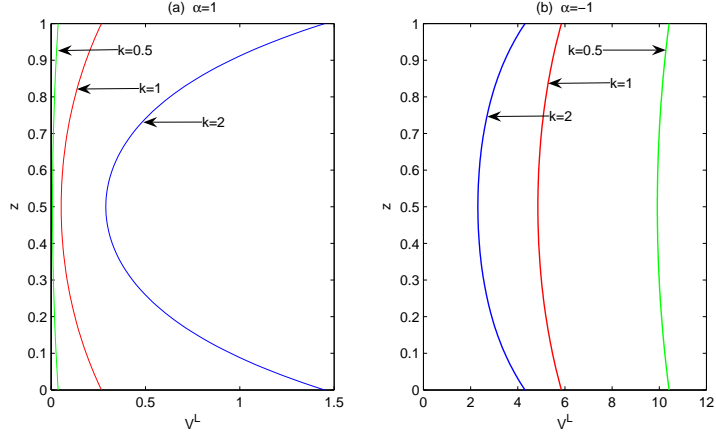}
\end{center}
\vskip -8mm
\caption{The averaged Lagrangian velocity profiles for $\beta=1$ (waves travelling in the same direction).}
\label{fig3}
\end{figure}

\begin{figure}
\begin{center}
\includegraphics*[height=6cm]{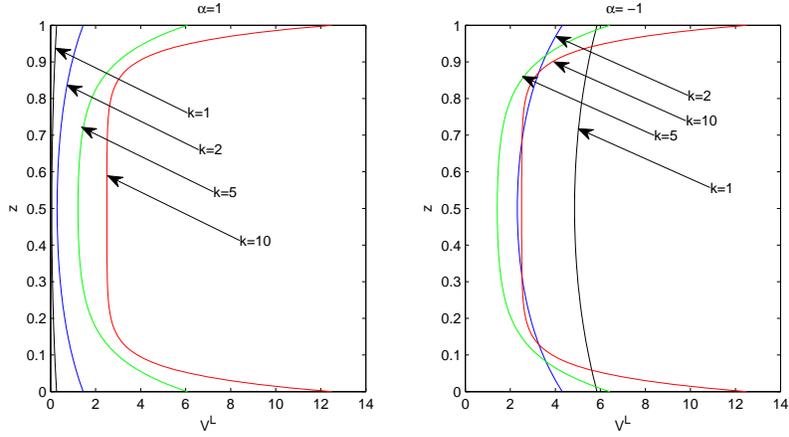}
\end{center}
\vskip -8mm
\caption{The averaged Lagrangian velocity profiles for $\beta=1$  (waves travelling in the same direction) and $k\geq 1$ (short waves).}
\label{fig4}
\end{figure}

\vskip 2mm
\noindent
The averaged Lagrangian velocity is given by
\[
\bar{\bv}_{1}^{L}=\bar{\bv}_{1}^{r}+\bV=V^{L}(z)\be_{x}, \quad V^{L}(z)=A(k)+V_{0}(z).
\]
Its profiles for various values of $k$ are shown in Fig. \ref{fig3}. In the short wave limit, $V^{L}(z)$ is constant
almost everywhere except narrow layers near the boundary where the velocity is much higher
(see Fig. \ref{fig4}). This can also be seen from the asymptotic formula
\[
V^{L}= k\left(\frac{1}{4}+e^{-2kz}+e^{-2k(1-z)}\right)+O(k e^{-k}), \quad k\to\infty.
\]

\vskip 2mm
\noindent
Now let $\beta=-1$. In this case,
boundary conditions (\ref{4.24})--(\ref{4.27}) take the form
\begin{equation}
\bar{u}_{1}^{r}\!\bigm\vert_{z=0}=B^{+}(k,x), \quad
\bar{u}_{1}^{r}\!\bigm\vert_{z=1}=B^{-}(k,x), \quad
\bar{w}_{1}^{r}\!\bigm\vert_{z=0}=C(k,x), \quad
\bar{w}_{1}^{r}\!\bigm\vert_{z=1}=C(k,x), \label{5.10}
\end{equation}
where
\begin{equation}
B^{\pm}(k,x)=\frac{k}{4}\left[\pm 1 + 6\alpha\frac{\cosh(k)}{\sinh^2(k)}\sin(2kx)\right],
\quad
C(k,x)=\alpha\frac{k\sin(2kx)}{\sinh(k)}. \label{5.11}
\end{equation}
Thus in the case of waves travelling in opposite directions, we have
nonzero boundary conditions
for the normal velocity in the averaged outer flow.

\vskip 2mm
\noindent
The Stokes drift velocity is given by
\begin{equation}
\bV(\br)=V_{1}(z)\be_{x}+V_{3}(x)\be_{z}, \quad
V_{1}(z)=-k\frac{\sinh[k(2z-1)]}{\sinh(k)}, \quad
V_{3}(z)=-C(k,x), \label{5.12}
\end{equation}
where $C(k,x)$ is given by (\ref{5.11}). Note that Eq. (\ref{5.12}) implies that
the normal component of the averaged Lagrangian velocity at the walls is zero: $w^{L}=\bar{w}_{1}^{r}+V_{3}=0$
at $z=0$ and $z=1$, which is in agreement with the general property discussed in Appendix A.

\vskip 2mm
\noindent
Further, we have
$\hat{\bom}_{1}^{r}=\Omega(x,z) \, \be_{y}$ where, as before,
$\Omega(x,z)=\pr_{z}u_{1}^{r}-\pr_{x}w_{1}^{r}$, and
Eq. (\ref{3.21}) can be written as
\begin{equation}
(\bar{\bv}^{r}_{1}\cdot\nabla)\bar{\bv}^{r}_{1}=
-\nabla \Pi_{3}+\nu\nabla^{2} \bar{\bv}^{r}_{1}+
\Omega \left[V_1(z) \, \be_{z}-V_3(x) \, \be_{x}\right]. \label{5.13}
\end{equation}
Thus, the averaged Eulerian velocity can be found by solving Eq. (\ref{5.13}) subject to
boundary conditions (\ref{5.10}). Then the averaged Lagrangian velocity is obtained by adding the Stokes drift
velocity given by Eq. (\ref{5.12}).

\begin{figure}
\begin{center}
\includegraphics*[height=8.5cm,width=18cm]{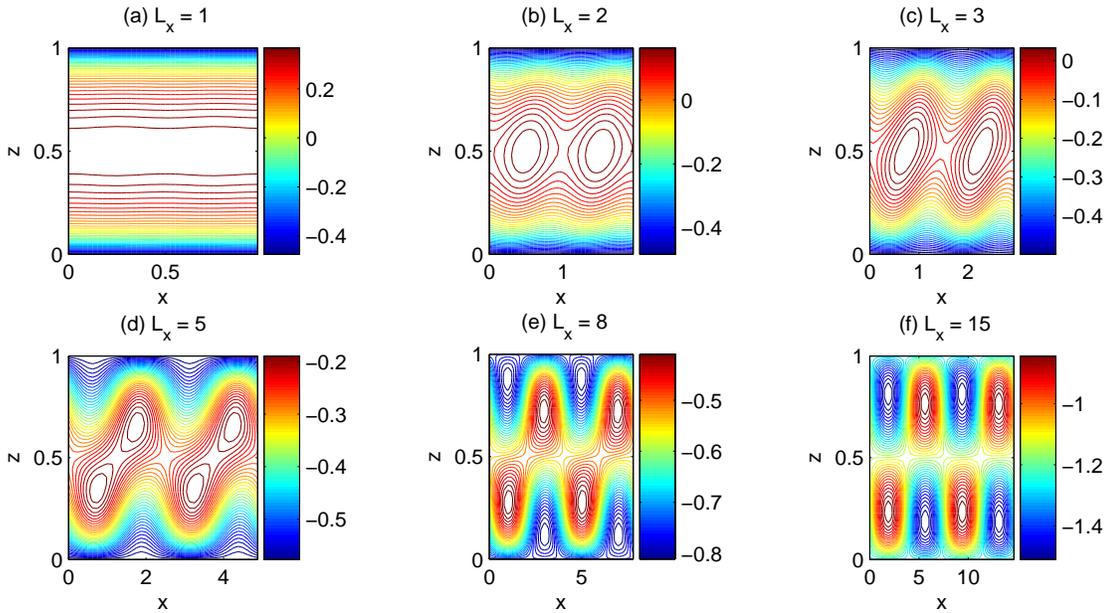}
\end{center}
\vskip -10mm
\caption{Streamlines of the averaged Lagrangian velocity for $\alpha=1$ and $\nu=1$ (waves travelling in opposite directions).}
\label{fig55}
\end{figure}

\vskip 0.2cm
\noindent
Equation (\ref{5.13}) supplemented with the incompressibility condition and boundary conditions (\ref{5.10})
were solved numerically. The streamlines of the averaged Lagrangian velocity $\bv^{L}$ for $\alpha=1$ and $\nu=1$ are shown in Fig. \ref{fig55}.
The corresponding values of $\max |\bar{\bv}_{1}^{r}|$ are given in Table 2. In the case $\alpha=-1$, the same flow pattern is  shifted by a quarter of the wavelength in the $x$ direction.
\begin{table}[h!]
\begin{center}
\begin{tabular}{|c|c|c|c|c|c|c|}
\hline
$L_x$  &1 &2 &3 &5 &8 &15 \\
\hline
$\max |\bar{\bv}_{1}^{r}|$  &0.80301 &0.60373   &0.71208 &1.0467  &1.6146  &2.9929  \\
\hline
\end{tabular}
\end{center}
  \vskip -3mm
  \caption{$\max |\bar{\bv}_{1}^{r}|$ for $\beta=-1$, $\alpha=1$ and $\nu=1$ (waves travelling in opposite directions).}
\end{table}


\subsection{General periodic vibrations}

\noindent
Here we will briefly describe how to treat
the case of general periodic (both in time and space) transverse vibrations of the wall.
We focus our attention on the travelling waves for which we will present an explicit solution.

\vskip 2mm
\noindent
Let $\bsig$ be the projection of $\bzeta$ (defined by Eq. (3.18)) onto the $xy$ plane, i.e.
\begin{equation}
\bsig = \bzeta - (\bzeta\cdot\be_{z})\be_{z},  \label{5.15}
\end{equation}
and let
\begin{equation}
\bsig^{a} = \bsig\!\bigm\vert_{z=0}, \quad
\bsig^{b} = \bsig\!\bigm\vert_{z=1}. \label{5.16}
\end{equation}
Let us express boundary conditions (\ref{4.24}),  (\ref{4.25}) in terms of
$f$, $g$, $\bsig^{a}$ and $\bsig^{b}$.

\vskip 2mm
\noindent
First, we note that since $\bsig^{a}$ and $\bsig^{b}$ are periodic functions of $\tau$ with period $2\pi$, they can be written as
Fourier series
\begin{equation}
\bsig = \sum_{n}\hat{\bsig}_{n}(\br,z)e^{in\tau}, \quad
\bsig^{a} = \sum_{n}\hat{\bsig}^{a}_{n}(\br)e^{in\tau}, \quad
\bsig^{b} = \sum_{n}\hat{\bsig}^{b}_{n}(\br)e^{in\tau}. \label{5.17}
\end{equation}
Here and in what follows (unless explicitly specified otherwise) summation is performed over all integers.
Also, in Eq. (\ref{5.17}),
\[
\hat{\bsig}_{0}(\br,z)\equiv 0, \quad \hat{\bsig}^{a}_{0}(\br)\equiv 0, \quad \hat{\bsig}^{b}_{0}(\br)\equiv 0.
\]
Now boundary conditions (\ref{4.3}) and (\ref{4.9}) can be written as
\begin{equation}
\bU^{a}_{0}\!\bigm\vert_{s=0}=-i \sum_{n} n \hat{\bsig}^{a}_{n}e^{in\tau}, \quad
\bU^{b}_{0}\!\bigm\vert_{q=0}=-i \sum_{n} n \hat{\bsig}^{b}_{n}e^{in\tau}.
\label{5.18}
\end{equation}
Periodic solutions of Eqs. (\ref{4.2}) and (\ref{4.8}) subject to boundary conditions (\ref{5.18})
and the condition of decay at infinity (in $s$ and $q$ respectively) are given by
\begin{eqnarray}
\bU^{a}_{0} =-i \sum_{n} n \hat{\bsig}^{a}_{n}e^{in\tau-\mu_{n}s}, \quad
\bU^{b}_{0} =-i \sum_{n} n \hat{\bsig}^{b}_{n}e^{in\tau-\mu_{n}q}, \label{5.19}
\end{eqnarray}
where
\begin{equation}
\mu_{n} = \sqrt{\frac{\vert n\vert}{2\nu}}(1+i \, \mathrm{sgn}(n)). \label{5.20}
\end{equation}
It is shown in Appendix B that boundary conditions  for $\bar{\bu}_{1}^{r}$ (given by (\ref{4.24}) and (\ref{4.25}))
can be written as
\begin{eqnarray}
&&\bar{\bu}^{r}_{1}\!\bigm\vert_{z=0}=
\overline{f_{\tau} \, \bsig_{z}}\!\bigm\vert_{z=0}
-\frac{3}{2}\overline{\bsig^{a}_{\tau}(\nabla_{\parallel}\cdot\bsig^{a})}
-\frac{1}{4}\nabla_{\parallel}\left( \sum_{n} \vert n\vert \vert\hat{\bsig}^{a}_{n}\vert^{2}\right)
-\sum_{n} \vert n\vert \,
\mathrm{Re}\left[\hat{\bsig}^{a*}_{n}(\nabla_{\parallel}\cdot\hat{\bsig}^{a}_{n}) \right],
\qquad\quad \label{5.21} \\
&&\bar{\bu}^{r}_{1}\!\bigm\vert_{z=1}=
\overline{g_{\tau} \, \bsig_{z}}\!\bigm\vert_{z=1}
-\frac{3}{2}\overline{\bsig^{b}_{\tau}(\nabla_{\parallel}\cdot\bsig^{b})}
-\frac{1}{4}\nabla_{\parallel}\left( \sum_{n} \vert n\vert \vert\hat{\bsig}^{b}_{n}\vert^{2}\right)
-\sum_{n} \vert n\vert \,
\mathrm{Re}\left[\hat{\bsig}^{b*}_{n}(\nabla_{\parallel}\cdot\hat{\bsig}^{b}_{n}) \right].
\qquad\quad \label{5.22}
\end{eqnarray}
From now on we will restrict our analysis to the two-dimensional case, where
both $f$ and $g$ do not depend on $y$ and
${\bu}^{r}_{1}=u^{r}_{1}\be_{x}$,  ${\bsig}=\sigma\be_{x}$. Equations (\ref{5.21}) and (\ref{5.22}) simplify to
\begin{eqnarray}
&&\bar{u}^{r}_{1}\!\bigm\vert_{z=0}=
\overline{f_{\tau} \, \sigma_{z}}\!\bigm\vert_{z=0}
-\frac{3}{2} \, \overline{\sigma^{a}_{\tau}\sigma^{a}_{x}}
-\frac{3}{4} \, \partial_{x}\left( \sum_{n} \vert n\vert \vert\hat{\sigma}^{a}_{n}\vert^{2}\right),
\qquad\quad \label{5.23} \\
&&\bar{u}^{r}_{1}\!\bigm\vert_{z=1}=
\overline{g_{\tau} \, \sigma_{z}}\!\bigm\vert_{z=1}
-\frac{3}{2} \, \overline{\sigma^{b}_{\tau}\sigma^{b}_{x}}
-\frac{3}{4} \, \partial_{x}\left( \sum_{n} \vert n\vert \vert\hat{\sigma}^{b}_{n}\vert^{2}\right).
\qquad\quad \label{5.24}
\end{eqnarray}
For simplicity, we consider only vibrations of the walls in the form of plane waves of arbitrary shape travelling
in the same direction\footnote{More general vibrations can be treated similarly.}, namely, we assume that
\begin{equation}
f(x,\tau)=F(k x -\tau), \quad  g(x,\tau)=\alpha F(k x -\tau), \label{5.25}
\end{equation}
where $F$ is an arbitrary $2\pi$-periodic function of a single variable $h=k x -\tau$; $k$ and $\alpha$
are the same as those defined in Section 5.2. Let $\hat{F}_{n}$ be the Fourier coefficients of $F$. Then the solution
of problem (\ref{3.5}) is
\begin{equation}
\phi_{0}(z,h)=\sum_{n} in b_{n}(z)\hat{F}_{n}e^{inh} \label{5.26}
\end{equation}
where
\[
b_{n}(z)=\frac{\cosh[\vert k n\vert (1-z)]-\alpha \cosh[\vert k n\vert z] }{\vert k n\vert \sinh\vert k n\vert }.
\]
Further, we have
\begin{equation}
\sigma(z,h)=-ik\sum_{n} n b_{n}(z)\hat{F}_{n}e^{inh} \label{5.27}
\end{equation}
and
\begin{equation}
\sigma^{a}(h)=i \alpha k\sum_{n} n Q_{n}\hat{F}_{n}e^{inh},
\quad \sigma^{b}(h)=-ik\sum_{n} n Q_{n}\hat{F}_{n}e^{inh},
\quad Q_{n}\equiv\frac{1-\alpha \cosh\vert k n\vert}{\vert k n\vert \sinh\vert k n\vert }. \label{5.28}
\end{equation}
It follows from (\ref{5.28}) that
\begin{equation}
\hat{\sigma}^{a}_{-n}(x)=ik n \alpha Q_{n}\hat{F}_{n}e^{inkx},
\quad \hat{\sigma}^{b}_{-n}(x)=-ik n Q_{n}\hat{F}_{n}e^{inkx}. \label{5.29}
\end{equation}
Substitution of (\ref{5.27})--(\ref{5.29}) into (\ref{5.23}), (\ref{5.24}) and tedious but elementary
calculations yield (cf. Eq. (\ref{5.6}))
\begin{eqnarray}
&&\bar{u}^{r}_{1}\!\bigm\vert_{z=0}=
\bar{u}^{r}_{1}\!\bigm\vert_{z=1}=A(k),  \nonumber \\
&&A(k)=k\sum_{n}n^2\vert \hat{F}_{n}\vert^2
\frac{5+\cosh^2\vert kn\vert - 6\alpha \cosh\vert kn\vert }{2\sinh^2\vert kn\vert} \nonumber \\
&&\quad\quad \ =k\sum_{n=1}^{\infty}n^2\vert \hat{F}_{n}\vert^2
\frac{5+\cosh^2( kn) - 6\alpha \cosh( kn) }{\sinh^2( kn)}. \label{5.30}
\end{eqnarray}
Boundary conditions for $\bar{w}_{1}^{r}$ (given by (\ref{4.26}), (\ref{4.27})) reduce to
\begin{equation}
\bar{w}_{1}^{r}\!\bigm\vert_{z=0}=0, \quad
\bar{w}_{1}^{r}\!\bigm\vert_{z=1}=0. \label{5.31}
\end{equation}
Similar calculations of the Stokes drift velocity yield
\[
\bV=V_{0}(z)\be_{x}
\]
where (cf. Eq. (\ref{5.7}))
\begin{eqnarray}
V_{0}&=&2k\sum_{n}n^2\vert \hat{F}_{n}\vert^2 \frac{\cosh\vert kn\vert - \alpha}{\sinh^2\vert kn\vert}
\, \cosh\left[\vert kn\vert(2z-1)\right] \nonumber \\
&=&4k\sum_{n=1}^{\infty}n^2\vert \hat{F}_{n}\vert^2 \frac{\cosh( kn) - \alpha}{\sinh^2( kn)}
\, \cosh\left[ kn (2z-1)\right]. \label{5.32}
\end{eqnarray}
Boundary conditions (\ref{5.30}) and (\ref{5.31}) results in a constant solution of
Eq. (\ref{5.8}): $\bv_{1}^{r}=A(k)\be_{x}$, which is a generalisation of the solution (\ref{5.9}) obtained in
Section 5.2. The averaged Lagrangian velocity is
\[
\bV^{L}=V^{L}\be_{x}, \quad V^{L}\equiv A+V_{0}(z).
\]
Let us compute the total volume flux of the fluid through the channel (per unit length in $y$ direction).
We have
\begin{equation}
Q=\int_{0}^{1}V^{L} \, dz = A + \int_{0}^{1}V_{0}(z) \, dz
=\sum_{n=1}^{\infty}n^2\vert \hat{F}_{n}\vert^2 \, R_{n}(k) \label{5.33}
\end{equation}
where
\begin{equation}
R_{n}= \frac{k}{\sinh^2( kn)}\left(
5+\cosh^2( kn) - 6\alpha \cosh( kn)+
4\left[\cosh( kn) - \alpha\right] \, \frac{\sinh( kn)}{kn}\right). \label{5.34}
\end{equation}

\begin{figure}
\begin{center}
\includegraphics*[height=8.5cm,width=14cm]{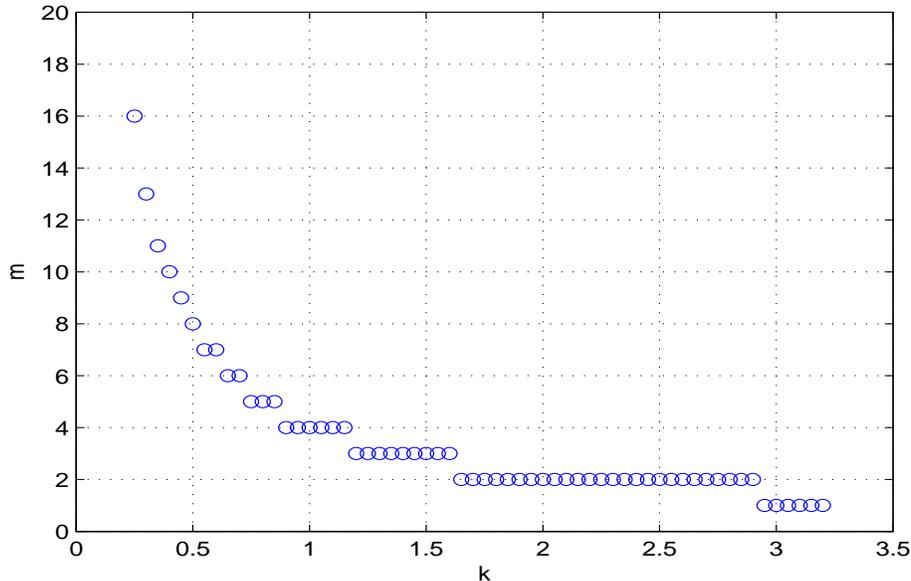}
\end{center}
\vskip -10mm
\caption{$m$, defined by (\ref{5.36}), as a function of $k$ for bending waves ($\alpha=1$).}
\label{Opt_flux}
\end{figure}

\vskip 2mm
\noindent
\textit{Optimal shape of the wave.}
Now we can answer the question on what shape of the travelling waves leads to the
the most efficient generation of the averaged flow. Mathematically, this can formulated as the following
optimization problem: maximize the total volume flux $Q$ on the set of all $\hat{F}_{n}$ such that
\begin{equation}
\sum_{n=1}^{\infty}n^2\vert \hat{F}_{n}\vert^2 =1 . \label{5.35}
\end{equation}
Let $m\in\mathbb{N}$ be such that
\begin{equation}
R_{m}=\max_{n\in\mathbb{N}} R_{n}. \label{5.36}
\end{equation}
It is not difficult to show that
$Q$ attains its maximum for
\[
\hat{F}_{n}=\frac{e^{i\lambda}}{\vert n\vert}  \, \delta_{\vert n\!\vert m} \quad (n=\pm 1,\pm 2,\dots),
\]
where $\lambda$ is any real number.
For contraction/expansion waves ($\alpha=-1$), the sequence $\{R_{n}\}$
is monotonically decreasing for all $k>0$, so that the maximum is attained at $n=1$. Thus, the most efficient
deformation of the walls is given by (cf. (\ref{5.5}))
\[
f(x,\tau)=2\cos(kx-\tau+\lambda), \quad g(x,\tau)=-f(x,\tau) .
\]
In the case of bending waves, the mode for which the maximum total
flux is attained depends on $k$ as shown in Fig. \ref{Opt_flux}.
One can see that $m$ is equal to 1 for $k\gtrsim 2.9$ (i.e. for sufficiently short waves) and increases
when $k$ decreases.
The most efficient
deformation has the form:
\[
f(x,\tau)=g(x,\tau)=2\cos[m(kx-\tau)+\lambda]/m .
\]

\setcounter{equation}{0}
\renewcommand{\theequation}{6.\arabic{equation}}

\section{Discussion}

\noindent
We have considered incompressible flows between two transversely vibrating solid walls
and proposed a general procedure for constructing an asymptotic expansion of solutions of the Navier-Stokes equations
in the limit when both the amplitude of vibrations and the thickness of the Stokes layer are small and have the same order
of magnitude. The procedure is based on the Vishik-Lyusternik method
and, in principle, can be used to construct as many terms of the expansion as necessary.
In the leading order, the averaged flow is described by the stationary Navier-Stokes equations
with an additional term which contains the leading-order Stokes drift velocity. In a slightly different context (for a flow
induced by an oscillating conservative body force), the same equations
had been derived earlier by Riley \cite{Riley2001}.

\vskip 0.2cm
\noindent
The general theory has been applied to two particular examples of steady streaming induced by
vibrations of the walls in the form of standing and travelling plane waves.
In the case of plane standing waves (bending or contraction/expansion waves),
the averaged flow has the form of a double array of vortices (see Fig. \ref{fig1a}). For short waves the vortices are concentrated near the walls,
while long waves produce vortices that fill the entire channel. The latter pattern is similar to flows
produced by long standing waves at small Reynolds numbers \cite{Carlsson}. For both bending and contraction/expansion waves, the flow patterns are qualitatively the same. However, the intensity of the averaged flow is different. Although it is of similar magnitude for both waves when the wavelength is small, for long waves, the maximum velocity decreases
much faster for bending waves
than for contraction/expansion waves, which is in agreement with the classical theory of peristaltic pumping
at low Reynolds numbers (see, e.g., \cite{Jaffrin}).

\vskip 0.2cm
\noindent
If vibrations of the wall have the form of plane harmonic waves which travel in the same direction
(this leads to a bending wave if the phase shift between these travelling waves is zero, and to a
contraction/expansion wave if the phase shift is $\pi$), the induced steady flow is a unidirectional
two-dimensional flow. The fortunate thing is that, in this case, the asymptotic equations can be solved analytically.
The Lagrangian velocity profile is symmetric relative to the channel central axis, with a minimum velocity
at the centre and maxima at the walls. In the short wave limit, the Lagrangian velocity profiles for both bending and contraction/expansion waves are the same: the velocity is nearly constant across the channel except for
narrow layers near the walls where velocity is much higher. For moderate and long waves, the velocity variation across
the channel is much smaller, and the magnitude of the velocity rapidly grows for the contraction/expansion wave
and decays for the bending wave when the wavelength increases.
This example may be viewed as an extension of the theory of peristaltic pumping to the case of high Reynolds numbers.

\vskip 0.2cm
\noindent
If vibrations of the walls have the form of plane harmonic waves traveling in opposite directions, the averaged
flow depends on the wavelength in a more complicated way. For short waves the averaged flow is a superposition of a shear flow with a linear velocity profile and a periodic array of weak vortices (`cat's eyes')
in the center of the channel. When the wavelength increases, the vortices first grow in size and magnitude. Then
each of them splits into a pair of vortices of the same sign, and at the same time new vortices appear near the walls and move
towards the centre of the channel. Eventually, for long waves, there are two arrays of alternating vortices
(similar to what we had in the case of standing waves).

\vskip 0.2cm
\noindent
As an example of general periodic vibrations, we have considered travelling contraction/expansion and bending
waves
of an arbitrary shape. Similarly to the case of a harmonic vibration, these produce unidirectional two-dimensional flows.
The averaged Lagrangian velocity is written as an infinite series in term of Fourier coefficients of the deformations
of the walls. The natural question that appears here is: what deformations of the walls lead to a maximum total
volume flux through the channel. It turns out  that the optimal shape always corresponds to a single harmonic.
For a contraction/expansion wave, it is always the first harmonic (with $m=1$).
For a bending wave, the optimal harmonic is uniquely determined by the parameter $k=2\pi/L_{x}$. For example,
if $k=1$, then $m=4$. It should be noted here that a contraction/expansion wave always produces a higher total volume flux
then a bending wave.

\vskip 0.2cm
\noindent
There are many open problems in this area. First, it is not quite clear how the present theory can be extended
to the case of high $R_{s}$. Second, we did not make any assumption about the characteristic length
scale of vibrations in the horizontal direction. Nevertheless, the examples show that the theory does not work when the horizontal length scale
is much smaller than the mean distance between the walls. In particular, Figures 2 and 5 shows that for $L_x< 1$ the flows  in Examples 1 and 2 become concentrated
near the walls. This suggests that in the limit of short waves, one can try to construct a theory with double boundary layers
(similar to what had been done by Stuart \cite{Stuart1966} and Riley \cite{Riley1965} for an oscillating cylinder at $R_{s}\gg 1$). Third,
it would be interesting to investigate the effect of an external mean flow (e.g., a flow produced by a mean pressure gradient)
on the steady steaming between two walls and vice versa. This problem has an additional parameter - the ratio of the characteristic velocity
of the mean flow to the amplitude of the velocity of the vibrating walls and, therefore,
one can expect a variety of different flow regimes. All these are problems for a future investigation.

\vskip 0.5cm
\noindent
{\bf Acknowledgments.} This work was initiated during a short visit of Andrey Morgulis to the University of York
under the University of York Research Development Visit scheme.


\setcounter{equation}{0}
\renewcommand{\theequation}{A.\arabic{equation}}

\section{Appendix A}

\noindent
Here we will show that
\begin{equation}
V_{3}\!\bigm\vert_{z=0}=-\nabla_{\|} \cdot
\overline{(f \nabla_{\|}\phi_{0})}\!\bigm\vert_{z=0}, \quad
V_{3}\!\bigm\vert_{z=1}=-\nabla_{\|} \cdot
\overline{(g \nabla_{\|}\phi_{0})}\!\bigm\vert_{z=1}, \label{A1}
\end{equation}
where $V_{3}$ is the z-component of the Stokes drift velocity $\bV$ given by Eq. (\ref{3.22}).

\vskip 2mm
\noindent
We have
\begin{eqnarray}
V_{3}&=& \bV\cdot\be_{z} =
\frac{1}{2}\be_{z}\cdot\overline{\left[\bzeta_{\tau},\bzeta\right]}=
-\overline{(\bzeta_{\tau}\cdot\nabla)(\bzeta\cdot\be_{z})}=
-\nabla\cdot\overline{[\bzeta_{\tau}(\bzeta\cdot\be_{z})]} \nonumber \\
&=&-\pr_{z}\overline{[(\bzeta_{\tau}\cdot\be_{z})(\bzeta\cdot\be_{z})]}
-\nabla_{\|}\cdot\overline{[\bu_{0}^{r}(\bzeta\cdot\be_{z})]}
=-\nabla_{\|}\cdot\overline{[\bu_{0}^{r}(\bzeta\cdot\be_{z})]}. \label{A2}
\end{eqnarray}
It follows from (\ref{3.17}) and (\ref{3.5}) that
$\bzeta\cdot\be_{z}\vert_{z=0}=f$
and
$\bzeta\cdot\be_{z}\vert_{z=1}=g$.
These and (\ref{A2}) imply the relations
\[
V_{3}\!\bigm\vert_{z=0}=-\nabla_{\|}\cdot\overline{(f\bu_{0}^{r})}\!\bigm\vert_{z=0} \quad
{\rm and} \quad
V_{3}\!\bigm\vert_{z=1}=-\nabla_{\|}\cdot\overline{(g\bu_{0}^{r})}\!\bigm\vert_{z=0}
\]
that are equivalent to (\ref{A1}).


\setcounter{equation}{0}
\renewcommand{\theequation}{B.\arabic{equation}}

\section{Appendix B}

\noindent
Here we will briefly describe the derivation of Eq. (\ref{5.21}). Equation (\ref{5.22}) can be derived
in exactly the same way.

\vskip 2mm
\noindent
First, we use (\ref{4.9}) and (\ref{4.11}) to rewrite Eq. (\ref{4.16}) as
\begin{eqnarray}
\bH^{a}_{1}&=&\left( s\nabla_{\parallel}^2\phi_{0}^{0}+W_{0}^{a}\!\bigm\vert_{s=0}-W_{0}^{a}
+(\bU_{0}^{a}-\bU_{0}^{a}\!\bigm\vert_{s=0})\cdot\nabla_{\parallel} f\right)\bU^{a}_{0s} \nonumber \\
&&- (\bU_{0}^{a}\cdot\nabla_{\parallel})\bU_{0}^{a}
- (\nabla_{\parallel}\phi_{0}^{0}\cdot\nabla_{\parallel})\bU_{0}^{a}
- (\bU_{0}^{a}\cdot\nabla_{\parallel})\nabla_{\parallel}\phi_{0}^{0}. \label{B.1}
\end{eqnarray}
Here $\phi_{0}^{0}=\phi_{0}\!\bigm\vert_{z=0}$. With the help of (\ref{4.5}), this can be further simplified to
\begin{eqnarray}
\bH^{a}_{1}&=& \left(s\nabla_{\parallel}^2\phi_{0}^{0} +
\nabla_{\parallel}\cdot\int\limits_{0}^{s}\bU^{a}_{0}(s')ds' \right)
\bU^{a}_{0s} \nonumber \\
&&- (\bU_{0}^{a}\cdot\nabla_{\parallel})\bU_{0}^{a}
- (\nabla_{\parallel}\phi_{0}^{0}\cdot\nabla_{\parallel})\bU_{0}^{a}
- (\bU_{0}^{a}\cdot\nabla_{\parallel})\nabla_{\parallel}\phi_{0}^{0}. \label{B.2}
\end{eqnarray}
Equation (\ref{4.2}), (\ref{4.3}) and the fact that $\bu_{0}^{r}=\nabla_{\parallel}\phi_{0}$ imply that
$\be_{z}\cdot\mathrm{curl} \bU_{0}^{a}=0$. Therefore, (\ref{B.2}) can be rewritten as
\begin{equation}
\bH^{a}_{1}= \left(s\nabla_{\parallel}^2\phi_{0}^{0} +
\nabla_{\parallel}\cdot\int\limits_{0}^{s}\bU^{a}_{0}(s')ds' \right)
\bU^{a}_{0s} - \nabla_{\parallel}
\left(\frac{\bU^{a2}_{0}}{2}+\bU^{a}_{0}\cdot\nabla_{\parallel}\phi_{0}^{0}\right). \label{B.3}
\end{equation}
It follows from (\ref{4.18}) that
\begin{equation}
\bar{\bU}^{a}_{1}\!\bigm\vert_{s=0}=
-\frac{1}{\nu}\int\limits_{0}^{\infty}\int\limits_{s'}^{\infty}
\bar{\bH}_{1}^{a}(s'') \, ds'' \, ds'=
-\frac{1}{\nu}\int\limits_{0}^{\infty}\bar{\bH}_{1}^{a}(s'')\int\limits_{0}^{s''}
 \, ds' \, ds'' =
 -\frac{1}{\nu}\int\limits_{0}^{\infty} s'' \, \bar{\bH}_{1}^{a}(s'') \, ds'' .  \label{B.4}
\end{equation}
On substituting (\ref{B.3}) into (\ref{B.4}), we obtain
\begin{eqnarray}
\bar{\bU}^{a}_{1}\!\bigm\vert_{s=0}&=&
\frac{2}{\nu} \, \overline{\nabla_{\parallel}^2\phi_{0}^{0}
\int\limits_{0}^{\infty} s \, \bU^{a}_{0} \, ds}
+\frac{1}{\nu} \nabla_{\parallel} \int\limits_{0}^{\infty}
\left(\overline{\frac{(\bU^{a}_{0})^2}{2}}+\overline{\bU^{a}_{0}\cdot\nabla_{\parallel}\phi_{0}^{0}}\right) s \, ds \nonumber \\
&&- \frac{1}{\nu}\overline{\int\limits_{0}^{\infty} s \, \bU^{a}_{0s}(s)
\left(\nabla_{\parallel}\cdot
\int\limits_{0}^{s}\bU^{a}_{0}(s') \, ds' \right)  ds}. \label{B.5}
\end{eqnarray}
It follows from (\ref{5.19}) that
\[
\int\limits_{0}^{\infty} s \, \bU^{a}_{0} \, ds= -i
\sum_{n}n \, \hat{\bsig}_{n}^{a} e^{in\tau}\int\limits_{0}^{\infty} s \, e^{-\mu_{n}s} \, ds
= -i
\sum_{n}n \, \hat{\bsig}_{n}^{a} \frac{1}{\mu_{n}^2} \, e^{in\tau}
=-\nu \sum_{n}\hat{\bsig}_{n}^{a}  \, e^{in\tau} = -\nu \bsig^{a}.
\]
Hence,
\begin{equation}
\frac{2}{\nu} \, \overline{\nabla_{\parallel}^2\phi_{0}^{0}
\int\limits_{0}^{\infty} s \, \bU^{a}_{0} \, ds}=-2 \,
\overline{\left(\nabla_{\parallel}\cdot\bsig^{a}_{\tau}\right)
\bsig^{a}}=2 \, \overline{\left(\nabla_{\parallel}\cdot\bsig^{a}\right)
\bsig^{a}_{\tau}}.  \label{B.6}
\end{equation}
Similar calculations yield
\begin{eqnarray}
&&\int\limits_{0}^{\infty}
\overline{\frac{(\bU^{a}_{0})^2}{2}} \, s\, ds =
\frac{\nu}{4}\sum_{n}\vert n\vert \vert \hat{\bsig}^{a}_{n}\vert^2, \quad
\overline{\nabla_{\parallel}\phi_{0}^{0}\cdot
\int\limits_{0}^{\infty}
\bU^{a}_{0} \, s\, ds} = -2\nu \, \overline{\bsig^{a}_{\tau}\cdot\bsig^{a} }=0, \label{B.7} \\
&&\overline{\int\limits_{0}^{\infty} s \, \bU^{a}_{0s}(s)
\left(\nabla_{\parallel}\cdot
\int\limits_{0}^{s}\bU^{a}_{0}(s') \, ds' \right)  ds}
=\nu\sum_{n}\vert n\vert \, \mathrm{Re}\left[\hat{\bsig}^{a*}_{n}
\left(\nabla_{\parallel}\cdot\hat{\bsig}^{a}_{n}\right)\right]
-\frac{\nu}{2} \, \overline{\left(\nabla_{\parallel}\cdot\bsig^{a}\right)\bsig^{a}_{\tau}}. \qquad \label{B.8}
\end{eqnarray}
Substituting (\ref{B.6})--(\ref{B.9}) into (\ref{B.5}), we find that
\begin{equation}
\bar{\bU}^{a}_{1}\!\bigm\vert_{s=0}=\frac{3}{2} \,
\overline{\left(\nabla_{\parallel}\cdot\bsig^{a}\right)\bsig^{a}_{\tau}}
+\frac{1}{4}\sum_{n}\vert n\vert \vert \hat{\bsig}^{a}_{n}\vert^2
+\sum_{n}\vert n\vert \, \mathrm{Re}\left[\hat{\bsig}^{a*}_{n}
\left(\nabla_{\parallel}\cdot\hat{\bsig}^{a}_{n}\right)\right].  \label{B.9}
\end{equation}
Finally, substitution of (\ref{B.9}) into (\ref{4.24}) results in boundary condition
(\ref{5.21}).

\end{document}